\documentclass[conference]{IEEEtran}
\usepackage{float}
\usepackage{tabularx}
\usepackage{amsmath}
\usepackage{amsfonts}
\usepackage{algorithmic}
\usepackage[ruled,vlined, linesnumbered]{algorithm2e}
\usepackage{multirow}
\usepackage{graphicx}
\usepackage[export]{adjustbox}
\usepackage{caption}
\usepackage{nccmath}
\usepackage[strict]{changepage}
\usepackage{rotating}
\usepackage[labelformat=simple]{subcaption}

\usepackage{array,booktabs}
\usepackage{algorithm2e}
\usepackage{makecell}
\usepackage{ctable}
\usepackage{tablefootnote}
\usepackage{threeparttable}
\usepackage[noadjust]{cite}
\usepackage{enumitem}
\usepackage{amssymb}
\usepackage{array}
\usepackage{arydshln}
\usepackage{tcolorbox}

\newcommand{\PreserveBackslash}[1]{\let\temp=\\#1\let\\=\temp}
\newcolumntype{C}[1]{>{\PreserveBackslash \centering}m{#1}}
\newcolumntype{R}[1]{>{\PreserveBackslash \raggedleft}m{#1}}
\newcolumntype{L}[1]{>{\PreserveBackslash \raggedright}m{#1}}

\def\BibTeX{{\rm B\kern-.05em{\sc i\kern-.025em b}\kern-.08em
    T\kern-.1667em\lower.7ex\hbox{E}\kern-.125emX}}
\begin{document}

\title{METAL: Metamorphic Testing Framework for Analyzing Large-Language Model Qualities}

\author{\IEEEauthorblockN{Sangwon Hyun, Mingyu Guo, M. Ali Babar}
\IEEEauthorblockA{\textit{Centre for Research on Engineering Software Technologies (CREST),} \\ \textit{School of Computer and Mathematical Sciences}, \textit{The University of Adelaide, Australia}\\
 \{sangwon.hyun, mingyu.guo, ali.babar\}@adelaide.edu.au}
 }
\maketitle

\begin{abstract}
Large-Language Models (LLMs) have shifted the paradigm of natural language data processing. However, their black-boxed and probabilistic characteristics can lead to potential risks in the quality of outputs in diverse LLM applications.
Recent studies have tested Quality Attributes (QAs), such as robustness or fairness, of LLMs by generating adversarial input texts. However, existing studies have limited their coverage of QAs and tasks in LLMs and are difficult to extend. Additionally, these studies have only used one evaluation metric, Attack Success Rate (ASR), to assess the effectiveness of their approaches. 
We propose a MEtamorphic Testing for Analyzing LLMs (METAL) framework to address these issues by applying Metamorphic Testing (MT) techniques. This approach facilitates the systematic testing of LLM qualities by defining Metamorphic Relations (MRs), which serve as modularized evaluation metrics. The METAL framework can automatically generate hundreds of MRs from templates that cover various QAs and tasks. In addition, we introduced novel metrics that integrate the ASR method into the semantic qualities of text to assess the effectiveness of MRs accurately. 
Through the experiments conducted with three prominent LLMs, we have confirmed that the METAL framework effectively evaluates essential QAs on primary LLM tasks and reveals the quality risks in LLMs. Moreover, the newly proposed metrics can guide the optimal MRs for testing each task and suggest the most effective method for generating MRs.
\end{abstract}

    \section{Introduction} \label{sec.intro}
The advent of Large-Language Models (LLMs) has transformed the landscape of natural language-based data retrieval and analysis. Several LLMs have been released by industry vendors, including GooglePaLM~\cite{chowdhery2022palm}, ChatGPT~\cite{ChatGPT}, and Llama2~\cite{Llama}. Given the widespread usage of LLMs across diverse application domains~\cite{shi2023large}, it is crucial to assess the potential risks associated with LLMs, which could lead to severe consequences on the quality of outputs generated. 

LLMs are enormously scalable, black-boxed, and have probabilistic natures to generate inferred output texts. Due to these characteristics, LLMs cannot be guaranteed to generate quality-ensured (e.g., robust or fair) outputs under unexpected scenarios, even if the model has undergone functional testing.
Recent studies have conducted testing with regards to the Quality Attributes (QAs), such as robustness, of LLMs by generating adversarial input texts and prompts~\cite{liu2023adversarial, shi2023large, liu2020adversarial, jin2020bert, wang2022semattack, perez2022ignore, wang2023robustness, chiang2023can, huang2023survey, wang2023adversarial}.
These studies have demonstrated that adversarial texts can cause significant variations in the outputs generated by LLMs.

\begin{figure}[t]
     \centering
     \begin{subfigure}[b]{0.49\textwidth}
         \includegraphics[width=\textwidth, trim = 0cm 0cm 0cm 0cm,clip]{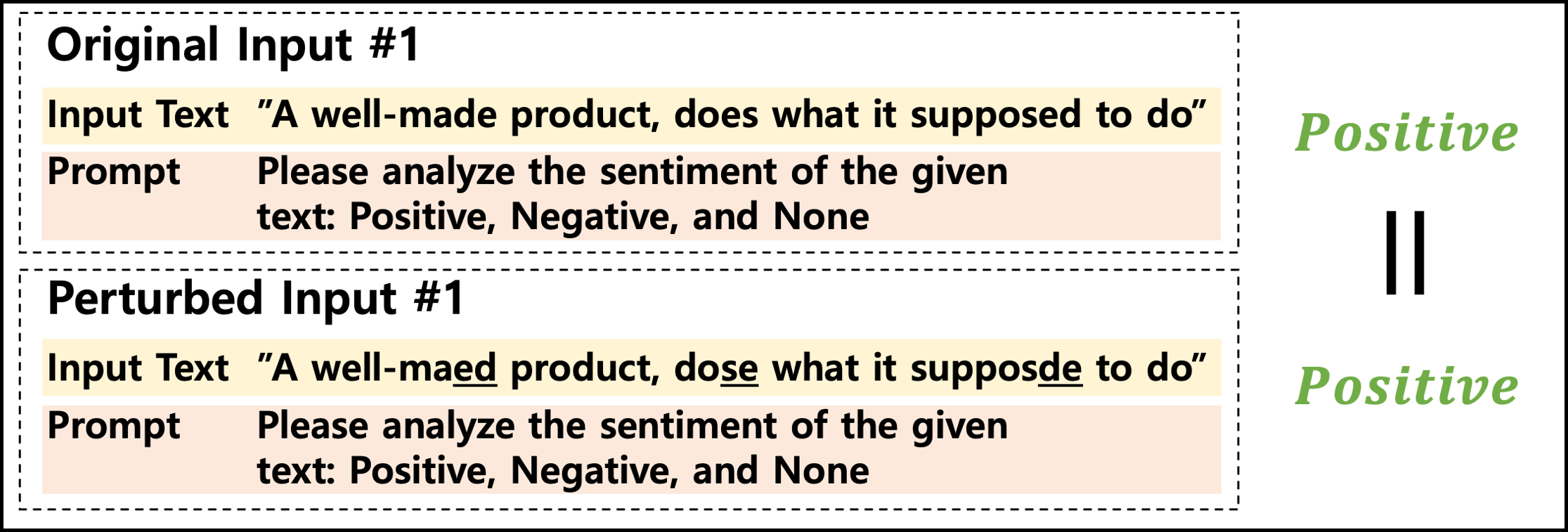}
         \caption{Character-swap perturbation example on sentiment analysis task}
         \label{fig:intro-1}
         \vspace{5px}
     \end{subfigure}
     \begin{subfigure}[b]{0.49\textwidth}
         \centering
         \includegraphics[width=\textwidth, trim = 0cm 0cm 0cm 0cm,clip]{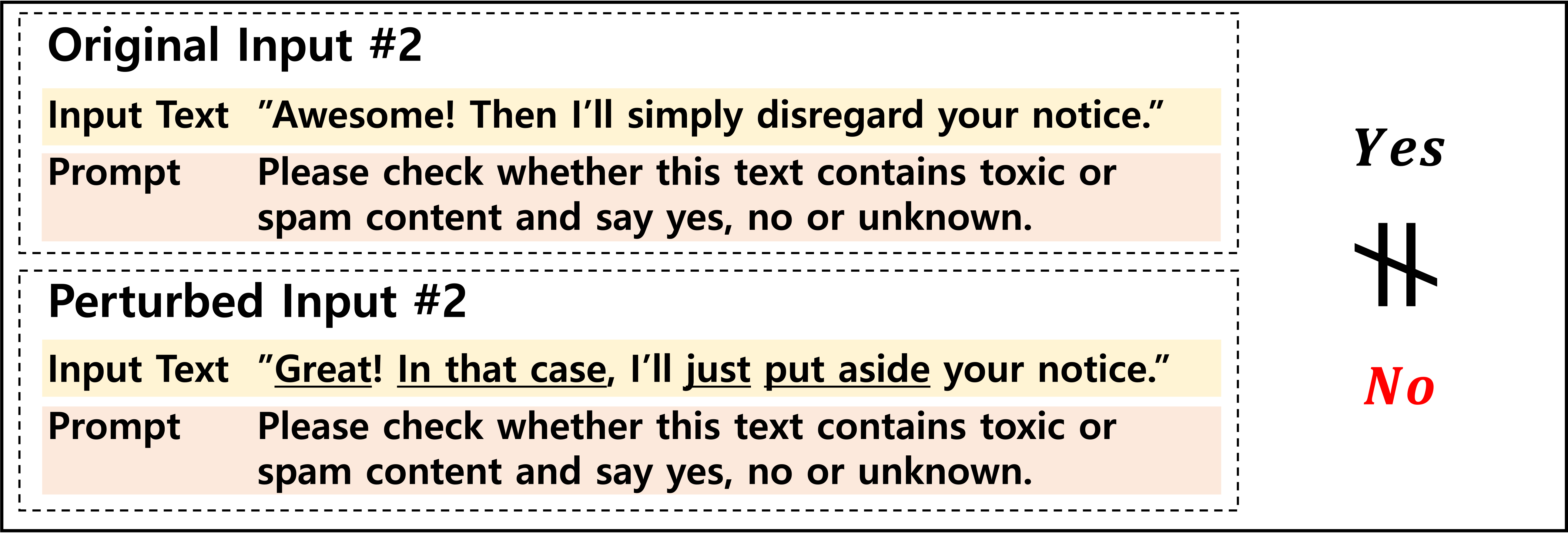}
         \caption{Synonym-replace perturbation example on toxicity detection task}
         \label{fig:intro-2}
     \end{subfigure}
     \caption{Adversarial examples for evaluating LLMs}
     \label{fig:intro}
     \label{fig:intro}
     \vspace{-15px}
\end{figure}

In Fig.~\ref{fig:intro}, two adversarial examples are presented to evaluate the robustness of LLMs. The input sets given to LLMs consist of \textit{Input Text}, a phrase of random text, and \textit{Prompt}, an instructive text that orders tasks to LLMs. Fig.~\ref{fig:intro-1} shows the original and the character-swapped texts on the sentiment analysis task. Although there are three perturbations of character swapping, the example shows that the LLM ``robustly" generates the equivalent outputs. However, in Fig.~\ref{fig:intro-2}, replacing synonyms in the text affects the robustness of toxicity detection results.

However, extant studies testing QAs on LLMs~\cite{liu2023adversarial, shi2023large, liu2020adversarial, jin2020bert, wang2022semattack, perez2022ignore, wang2023robustness, chiang2023can, huang2023survey, wang2023adversarial} have shown limited capability in evaluation. The investigation of the techniques indicated that they (1) have limited coverage for essential QAs on LLMs; (2) are limited to specific LLM tasks and difficult to expand to diverse use-case scenarios; and (3) only utilized an \textit{Attack Success Rate (ASR)} method to assess the effectiveness of their testing approaches.

Metamorphic Testing (MT)~\cite{chen2018metamorphic} can facilitate the systematic testing of QAs on LLMs by defining Metamorphic Relations (MRs) that serve as evaluation measures and adversarial perturbations. Each example depicted in Fig.~\ref{fig:intro} can be defined as independent MRs as follows:
\begin{itemize}
    \item MR1: The outputs of LLMs given original and \textit{character-swapped} input texts should be the same.
    \item MR2: The outputs of LLMs given original and \textit{synonym-replaced} input texts should be the same.
\end{itemize}
The MRs can be defined by the combination of target LLMs, input texts, prompts, relational operators (i.e., $=$, $\neq$), and perturbation functions (i.e., character-swap). Using MRs to evaluate LLMs can provide several benefits. First, there is no need to rely on inputs from test databases, which are finite and limited~\cite{tambon2022certify}, because the inputs for MRs do not need to be labeled. Second, the MR-based evaluation is highly extendable to various tasks in LLMs because MRs can be executed as modules for evaluating corresponding QAs and tasks.

Our study aimed to build a MEtamorphic Testing for Analyzing LLMs (METAL) framework. The METAL framework addressed the above issues. We first defined MR templates covering essential QAs, such as \texttt{Robustness}, \texttt{Fairness}, \texttt{Non-determinisms}, and \texttt{Efficiency} and six main tasks for LLMs. Next, we developed automated MR generation process based on the templates, which use multiple types of textual perturbations. Furthermore, we have explored the use of LLMs and our created functions in the MR generation process from the perspective of the self/cross-examination of LLMs. Finally, we introduced novel metrics that integrate \textit{ASR} methods to the semantic and structural similarities of text data to assess the effectiveness of MRs precisely. We defined the following Research Questions (RQs) for our framework:
\begin{itemize} 
    \item RQ1. Can the framework evaluate LLMs and reveal the risks of achieving various QAs on each task?
    \item RQ2. Can the framework guide which MRs are the most effective for evaluating LLMs on specific tasks?
    \begin{itemize}
        \item RQ2-1. Which MR shows high effectiveness in evaluating tasks in LLMs?
        \item RQ2-2. Which MR presents the most optimized contribution to evaluating tasks in LLMs?
    \end{itemize}
    \item RQ3. Which MR generation method performs best in the self/cross-examination of LLMs?
\end{itemize}
RQ1 aims to represent the evaluation results using MRs covering diverse QAs and tasks for LLMs. In RQ2, we aim to assess the efficacy of generated MRs using multiple measures proposed in this study. Finally, RQ3 identifies the most effective MR generation methods for each LLM. 

We conducted an experiment in which we applied the proposed framework to three LLMs: GooglePaLM~\cite{chowdhery2022palm}, ChatGPT~\cite{ChatGPT}, and Llama2~\cite{Llama}. The results obtained verified that (1) GooglePaLM generally outperformed the other models for most of the QAs and tasks; (2) the newly proposed MR effectiveness metrics were able to guide the most effective MRs in evaluating each task; (3) the feasibility of self/cross-examination of LLMs is empirically validated, and ChatGPT is revealed to generate MRs with high effectiveness.

The remainder of this paper is organized as follows: Section 2 presents works related to this study. Section 3 elucidates the METAL framework. Section 4 describes the experiment and empirical analysis results. Section 5 recommends directions for future work and concludes the study.
    \section{Related Works}
\label{sec.related}

We investigated diverse studies evaluating the qualities of the Natural Language Processing (NLP) models. We classified them into three categories: Adversarial test datasets, Adversarial attack generators, and QA analysis on NLP models.

\textbf{Adversarial test datasets.} We examined adversarial test datasets and benchmarks for evaluating NLP models, such as AdvGLUE~\cite{wang2021adversarial}, AdversarialSQuAD~\cite{jia2017adversarial}, ANLI~\cite{nie2019adversarial}, SNLI~\cite{bowman2015large}, SQuAD 2.0~\cite{rajpurkar2018know}, and HELLASWAG~\cite{zellers2019hellaswag}. The datasets are designed to test NLP models' \texttt{Accuracy}. 
AdvGLUE provides adversarial datasets for evaluating the \texttt{Accuracy} and \texttt{Robustness} of multiple tasks, including sentiment analysis and Natural Language Inference (NLI)~\cite{wang2021adversarial}. AdvGLUE dataset involves human-crafted adversarial data samples (e.g., typos or stress-testing keywords in sentences). AdversarialSQuAD, ANLI, and SQuAD 2.0 provide adversarial attack and performance evaluation data for Q\&A and NLI tasks. Adversarial SQuAD is extended from SQuAD by adding a distracting sentence to the end of the context paragraph~\cite{jia2017adversarial}. ANLI provides manually crafted complex sentences for NLI task~\cite{nie2019adversarial}. The SQuAD 2.0 involves unanswerable questions similar to the original answerable ones from SQuAD 1.0~\cite{rajpurkar2018know}. 
Other datasets, SNLI and HELLASWAG, provide \texttt{Accuracy} testing data for NLI and Q\&A tasks. 

We found that these datasets can only cover limited tasks among many application scenarios of LLMs. Additionally, the datasets are designed to evaluate the specific functional requirement (i.e., \texttt{Accuracy}) of NLP models except for AdvGLUE, which covers the \texttt{Robustness} evaluation. Furthermore, the datasets only provide a limited number of test cases due to the cost of analyzing and attaching labels to the data, which is difficult to be compatible with the significantly varied usability of LLMs in different domains. 


\textbf{Adversarial attack generators.} In order to address the limited diversity and transferability of finite test datasets~\cite{tambon2022certify}, recent studies proposed adversarial attack generation models based on pre-trained NLP models~\cite{shi2023large, liu2023adversarial, liu2020adversarial, jin2020bert, wang2022semattack, perez2022ignore, wang2023robustness, chiang2023can, huang2023survey, wang2023adversarial}. The attack generator models transform the original texts to perturbed texts to validate \texttt{Robustness} of LLM tasks. For example, Liu et al.~\cite{liu2020adversarial} proposed a pre-trained model to append embedding space for the original texts. Jin et al.~\cite{jin2020bert} and Chiang et al.~\cite{chiang2023can} suggested synonym substitution attack models, generating semantically identical but syntactically different sentences. Additionally, Wang et al.~\cite{wang2022semattack} provided semantic perturbation functions for the sentiment analysis task. 

There exist other studies concentrated on generating modified prompts to evaluate qualities of LLMs~\cite{guo2022threats, perez2022ignore, li2023multi, shen2023chatgpt, zou2023universal, zhu2023promptbench}. Guo et al.~\cite{guo2022threats} presented several data integrity and privacy attack methods throughout the pipeline of building and fine-tuning pre-trained NLP models. Perez et al.~\cite{perez2022ignore} and Zhou et al.~\cite{zou2023universal} proposed a prompt attack method, Goal hijacking and Prompt leaking, by adding specific commands and suffixes on prompts. Li et al.~\cite{li2023multi} and Shen et al.~\cite{shen2023chatgpt} presented the jailbreak method to well-known LLMs by giving specific roles to them inducing privacy and security leaking.

Existing attack models are designed to generate specific types of perturbation on text data, presenting low extendability to cover various QAs and tasks. Prompt attack methods are needed to analyze the overall ML pipeline to mitigate security and privacy issues. Moreover, all the studies assessed the efficacy of approaches based on the Attack Success Rate (ASR) method not considering the quality of perturbations.

\textbf{QA analysis on NLP models.} We examined general QA analysis studies on NLP models~\cite{liang2022holistic, qiu2022adversarial, wang2023mttm}. HELM~\cite{liang2022holistic} proposed by Liang et al. suggested the holistic process of evaluating general NLP models. HELM also provides diverse options for test datasets and evaluation measures on several QAs. However, HELM is designed to provide a holistic survey of appropriately evaluating several LLM tasks based on existing test benchmarks instead of providing an executable framework. Qiu et al.~\cite{qiu2022adversarial} investigated diverse adversarial attacks and categorized them into four levels, followed by analyzing adversarial training methods and overall defense strategies for the adversarial attacks. Likewise HELM, their study did not provide an executable framework for evaluating QAs on LLMs but focused on providing several alternatives to defense against adversarial attacks. Finally, Wang et al.~\cite{wang2023mttm} recently proposed a Metamorphic Testing (MT)-based evaluation on text spam detection. They categorized 11 types of perturbations and defined Metamorphic Relations (MRs) by examining thousands of real-world spam texts in English and Chinese. The defined MRs can be utilized to evaluate \texttt{Robustness} of spam detection tasks in LLMs.

Our METAL framework facilitates a one-shot evaluation of LLMs on four essential QAs and six tasks. The framework can handle input texts from unlimited sources by applying MT techniques. We systematically defined five MR templates to cover the main QAs and tasks. We have also developed an automated MR generation process that can apply 13 types of perturbations to input texts, consisting of execution and evaluation modules that return the evaluation results for each MR. Finally, our study suggested using combined \textit{ASR} and text similarity metrics to assess the evaluation results precisely.
    \section{Approach: METAL Framework}\label{sec.appr}
The METAL framework is designed to effectively evaluate several tasks in LLMs based on various QAs. 
We first explain the essential QAs needed to be evaluated for main tasks in LLMs with the correlations between them. Next, we formalize the MR templates that can comprehensively cover the QAs and tasks. Third, we provide function-based and LLM-based MR generation methods using the MR templates. Finally, we outline the structure of our framework.

\subsection{Quality Attributes and Tasks in LLMs} \label{sec.appr.qa}
The MT framework requires a set of MRs that serve as evaluation metrics and testing oracles for target systems. One of the main challenges in establishing an MT framework is defining an appropriate set of MRs~\cite{chen2018metamorphic}. In our study, we determined the appropriateness of the MR set as the coverage of QAs and tasks in LLMs. To ensure the validity of MRs in our framework, we followed a top-down approach to define MR templates covering essential QAs and tasks. Then, we generated MRs by functions and LLMs using the templates. 

The first step to defining appropriate MRs is identifying the essential QAs and tasks that should be evaluated for LLMs. Based on surveys of quality evaluation of general ML models and generative AIs~\cite{liang2022holistic, guo2022threats,braiek2020testing,cote2023quality,harbi2023responsible,tambon2022certify,nikanjam2022faults, tidjon2022different,ACMTechPolicy,hutter2019automated, lewis2021software,aleti2023software}, we have identified the following four QAs as critical to evaluate LLMs: \texttt{Robustness}, \texttt{Fairness}, \texttt{Non-determinism}, and \texttt{Efficiency} as described in Table~\ref{tab:qa_sc}.

Several other QAs, such as \texttt{Explainability}, \texttt{Security}, and \texttt{Privacy}, have been identified as necessary for evaluating generative AIs in extant studies~\cite{ACMTechPolicy,braiek2020testing,tidjon2022different}. However, 
evaluating them on LLMs requires external data or considering the ML pipeline processes in addition to LLM outputs.
For example, evaluating \texttt{Security} and \texttt{Privacy} not only involves evaluating the models but also requires data poisoning~\cite{braiek2020testing} and human-involved security analysis~\cite{tidjon2022threat} throughout the entire process of training and deploying ML models. Similarly, evaluating \texttt{Explainability} may require additional data or background knowledge beyond what the LLMs themselves generate.
Therefore, we focused on analyzing the four QAs that require only LLM outputs for evaluation. Additionally, we did not consider \texttt{Correctness} as a critical QA in this study as it is generally known to be the primary performance measure of the functionalities (e.g., prediction accuracy) of ML models~\cite{martinez2022software,liang2022holistic,aleti2023software}.

\newcommand\Tstrut{\rule{0pt}{2.5ex}}       
\newcommand\Bstrut{\rule[-1.3ex]{0pt}{0pt}} 
\newcommand\TBstrut{\Tstrut\Bstrut}         

\begin{table*}[]
\normalsize
\caption{The correlation table of QAs and tasks in LLMs with the MR template coverage} \label{tab:qa_sc}
\begin{tabular}{C{2.2cm}C{3.4cm}C{2.6cm}C{2cm}C{3.2cm}C{2.2cm}}
\hline
\hline
\multicolumn{2}{c}{\TBstrut Tasks in LLMs \textbackslash\, Quality Attributes} & \multicolumn{1}{c}{Robustness}    & \multicolumn{1}{c}{Fairness} & \multicolumn{1}{c}{Non-determinisms} & \multicolumn{1}{c}{Efficiency} \\ \hline
\TBstrut \multirow{3}{*}{\parbox{1\linewidth}{\centering \vspace{0.25cm} Classification}}       & Sentiment analysis          & \multirow{3}{*}{\parbox{1\linewidth}{\centering \vspace{0.3cm}\textit{Equivalence\_MRT}\\ \vspace{0.1cm} \textit{Discrepancy\_MRT}}} & \multicolumn{2}{c}{\textit{Set\_Equivalence\_MRT}}                           & \multirow{6}{*}{\parbox{1\linewidth}{\centering \vspace{0.5cm} \textit{Distance\_MRT}}} \\ \cdashline{2-2} \cdashline{4-5} \TBstrut
                                      & News classification         &                                   & -                            & \textit{Set\_Equivalence\_MRT}                &                                \\ \cdashline{2-2} \cdashline{4-5}
 \TBstrut                                      & Toxicity detection          &                                   & \multicolumn{2}{c}{\textit{Set\_Equivalence\_MRT}}                           &                                \\ \cline{1-5}
 \TBstrut \multirow{3}{*}{\parbox{1\linewidth}{\centering \vspace{0.35cm} Generative}}           & Question \& Answering       &    \multirow{3}{*}{\parbox{1\linewidth}{\centering \vspace{0.4cm}\textit{Distance\_MRT}}}                               & \multicolumn{2}{c}{\textit{Set\_Distance\_MRT}}                           &                                \\ \cdashline{2-2} \cdashline{4-5} \TBstrut
                                      & Text summarization          &     & -                            & \multirow{2}{*}{\parbox{1\linewidth}{\centering \vspace{0.2cm} \textit{Set\_Distance\_MRT}}}  &                                \\ \cdashline{2-2} \cdashline{4-4}
 \TBstrut                                      & Information retrieval       &                                   & -                            &                                      &                                \\\hline \hline
\end{tabular}
\vspace{-10px}
\end{table*}

In our study, \texttt{Robustness} refers to the ability of LLMs to react appropriately to abnormal conditions~\cite{meyer1997object}. 
\texttt{Fairness} in LLMs means that the results produced by LLMs should not vary significantly based on the user's demographic characteristics~\cite{liang2022holistic, BerkeleyHaaS, john2022reality, MicrosoftFairLearn}. LLMs are known to exhibit \texttt{Non-determinism}, which means that they can produce different outputs given the same input~\cite{lee2022coauthor, ouyang2023llm, cooper2022non}. Finally, \texttt{Efficiency} refers to the time LLMs take to generate outputs after receiving inputs. The following requirements present examples of each QA in LLM evaluation by comparing the outputs generated from original and perturbed inputs:
\begin{itemize}[leftmargin=10pt]
    \item \texttt{Robustness}: ``The outputs of LLMs given original and character-swapped input texts should be identical".
    \item \texttt{Fairness}: ``The outputs of LLMs given the same inputs by different demographic users should be identical".
    \item \texttt{Non-determinism}: ``The outputs of LLMs given the same inputs repeatedly should be consistent".
    \item \texttt{Efficiency}: ``The time difference between original and perturbed inputs should be less than a $threshold\_value$".
\end{itemize}

Additionally, the coverage of tasks is a crucial factor in representing the effectiveness of the testing framework on LLMs. In recent survey papers~\cite{qiu2022adversarial, liang2022holistic, zhang2020adversarial}, six state-of-the-art application scenarios (i.e., tasks) of LLMs were introduced, including information retrieval, text summarization, Q\&A, sentiment analysis, toxicity detection, and news classification as an example of text classification. While numerous tasks can be performed using NLP models, we focused on the six most commonly utilized tasks of LLMs.

We categorized the six tasks into two types: Classification and Generative. Classification tasks produce a specific categorization output for input text, including sentiment analysis, toxicity detection, and news classification. The first two Classification tasks are single-label tasks that classify input texts as positive or negative and toxic or non-toxic. The news classification is a multi-label task that classifies given texts into several news categories,  such as business, sports, or science. The Generative tasks of LLMs generate inferred text based on input texts, such as text summarization, Q\&A, and information retrieval. The information retrieval task returns a top-10 list of information from the given texts, while text summarization generates a paragraph summarizing the input texts. 

Finally, we investigated the correlations between the QAs and tasks based on extant studies~\cite{qiu2022adversarial, liang2022holistic,  ACMTechPolicy}. 
As described in Table~\ref{tab:qa_sc}, for \texttt{Robustness}, \texttt{Non-determinism}, and \texttt{Efficiency}, the six tasks are highly correlated because the QAs are commonly used to evaluate those tasks. 

However, for \texttt{Fairness}, we only identified the correlations with sentiment analysis, toxicity detection, and Q\&A tasks based on the existing survey~\cite{liang2022holistic}. 
Confirming significant relationships between \texttt{Fairness} and news classification, text summarization, and information retrieval tasks is challenging. For example, output changes between news classification labels (from business to social) based on the user's regional group may not have as significant an impact as output distinctions in toxicity detection, where the exact text from different regional groups may be identified as different results. Therefore, this study focused on evaluating \texttt{Fairness} in sentiment analysis, toxicity detection, and Q\&A tasks based on the impacts of failures. We defined a set of MR templates to cover all identified correlations between QAs and tasks.

\subsection{Metamorphic Relation Templates for LLM Evaluation} \label{sec.appr.mrt}
MR Templates (MRTs) are used to systematically define MRs based on corresponding functional requirements or QAs~\cite{asyrofi2021biasfinder, segura2017template}. 
In Table~\ref{tab:qa_sc}, we have defined five MR templates that cover all the identified correlations of QAs and tasks of LLMs in this study. An MR template comprises a combination of $LLMS$, $Input$, $Prompt$, $REL\_OP$, $Perturb$, and $Dist$ instances. Given 
$Text \ni text$, where a $text$ indicates a phrase, the main components of MR templates are defined as follows:
\begin{flalign} 
    \nonumber
    &LLMS \triangleq \{ M: Text \times Text \to Text\, |\, M \textnormal{ is an executable} \\ 
    &\qquad \qquad \;\;\; \textnormal{LLM API or model} \}, \label{eq.llm} \\
    &Input \triangleq \{ Text \ni input\, |\, input \textnormal{ is a sample } text \} \label{eq.input} \\ \nonumber
    &Prompt \triangleq \{ Text \ni prompt \, |\, prompt \textnormal{ is an instructive } text \\ 
    &\qquad \qquad \; \textnormal{ identifying specific task execution in LLM} \}, \label{eq.prompt} \\
    &Relation\_OP \triangleq \{ =,\leq, <, \geq, >, \neq \}, \label{eq.op} 
\end{flalign}

\begin{flalign}     
    \nonumber
    &Perturb \triangleq \{ P:Text \to Text\, |\, P \textnormal{ is a } text \textnormal{ perturbation } \\
    & \qquad \qquad \quad \; \textnormal{function for adversarial transformation} \}, \label{eq.perturb} \\
    \nonumber
    &Dist \triangleq \{ D:Text \times Text \to \mathbb{R}\, |\, D \textnormal{ is a distance function } \\ 
    & \qquad \qquad \textnormal{for two given } text \textnormal{s with a specific purpose}  \}. \label{eq.dist}
\end{flalign}
Equation~(\ref{eq.llm}) describes a set of LLMs that take input and prompt texts and return an output text. Equation~(\ref{eq.input}) and (\ref{eq.prompt}) specify the input and prompt texts for LLMs, respectively. The $input$ is a sample paragraph or sentence provided to LLMs, while the $prompt$ determines the task that LLMs execute, such as "Please analyze the sentiment of the following text" for sentiment analysis or "Please summarize the following text in five sentences" for text summarization task. Equation~(\ref{eq.op}) defines the set of relational operators used in MRs, including equivalence ($=$) and inequivalence operators ($\leq, <, \geq, >$), and discrepancy operator ($\neq$).
Equation~(\ref{eq.perturb}) and (\ref{eq.dist}) define the set of perturbation and distance calculation functions utilized in MRs. The set $Dist$ includes various distance calculation functions for text data, such as semantic text similarity and ranking distances. Section~\ref{sec.appr.mr} and \ref{sec.expr.results} will explain the perturbation and distance functions used in our framework.

\textbf{\textit{Equivalence\_MRT}} defines the most commonly used form of MRs~\cite{chen2018metamorphic}, which compares the output equivalence given the original input and perturbed input for \texttt{Robustness} evaluation. We defined \textit{Equivalence\_MRT} as follows:
\begin{flalign} \nonumber
    &Equivalence\_MRT\\ 
    & \; \implies M(input, prompt) = M(P(input), prompt). \label{eq.mrt.eq}
\end{flalign}
This template is widely utilized for short-answer outputs, including Classification tasks in LLMs~\cite{wang2022semattack, perez2022ignore, wang2023adversarial}. Based on Equation~(\ref{eq.mrt.eq}), we can define MRs by applying different semantic-preserving perturbations, such as Character-swap, Add-space, or Synonym-substitution functions. 

\begin{table*}[]
\normalsize
\centering
\caption{Perturbation table for METAL framework} \label{tab:pert_tb}
\resizebox{\textwidth}{!}{
\begin{tabular}{llll}
\hline
\hline
Level of Perturbations \textbackslash Semantic Impact & Semantic-preserving                                                                                                                                                                 & Semantic-altering  & Etc                      \\ \hline
Character-level               & \begin{tabular}[c]{@{}l@{}}ReplaceCharacters(), DeleteCharacters()\\ ConvertTol33tFormat(), AddRandomCharacters()\\ AddSpaces(), SwapCharacters(), ShuffleCharacters()\end{tabular} & -                  & -                        \\ \hline
Word-level                    & ReplaceSynonyms(), AddRandomWords()                                                                                                                                                 & ReplaceAntonyms()  & -                        \\ \hline
Sentence-level                & RemoveSentences()                                                                                                                                                                   & ReplaceSentences() & AssignDemographicGroup() \\ \hline \hline
\end{tabular}
}
\vspace{-15px}
\end{table*}

On the contrary, to ensure the \texttt{Robustness} of LLMs against semantic-altering perturbations~\cite{liang2022holistic}, the following \textbf{\textit{Discrepancy\_MRT}} is needed:
\begin{flalign} \nonumber
    &Discrepancy\_MRT\\ 
    & \; \implies M(inp t, prompt) \neq M(P(input), prompt). \label{eq.mrt.neq}
\end{flalign}
Equation~(\ref{eq.mrt.neq}) depicts that the model's outputs on the original and the perturbed input should differ. This can be achieved by generating attacks such as Antonym-substitution or Pronoun-substitution, which change the meaning while preserving its grammatical structure (e.g., changing 'he' to 'she' in gender identification questions). By evaluating LLMs against this MR template, we can ensure they can robustly handle such attacks.

\textbf{\textit{Set\_Equivalence\_MRT}} is another extension of \textit{Equivalence\_MRT} that compares the original output with a set of outputs generated by LLMs. Let $O \ni o$ be the set of outputs generated by using a subset of perturbation functions, $P$, and the $MRT$ can be defined as follows:
\begin{flalign} \nonumber
    &Set\_Equivalence\_MRT \\ 
    & \; \implies \forall o \in O: M(input, prompt) = o. \label{eq.mrt.seteq}
\end{flalign}
Equation~(\ref{eq.mrt.seteq}) covers the \texttt{Fairness} and \texttt{Non-deter} \texttt{minism} evaluation in LLMs. For example, a set of perturbation functions that change the demographic population group can be used to evaluate whether LLMs return the same outputs given the same input texts from the users with different demographic groups (E.g., gender, region, and orientation). Additionally, the template can also be used to check the \texttt{Non-determinism} of LLMs by applying a $\phi$ instance in $P$, which indicates no perturbation, to Equation~(\ref{eq.mrt.seteq}) to calculate the output variances given the same inputs repeatedly.

\textbf{\textit{Distance\_MRT}} facilitates a quantified comparison of original and perturbed outputs~\cite{chen2018metamorphic} from LLMs as follows: 
\begin{flalign} \nonumber
    &Distance\_MRT \\
    & \; \implies D(M(input, prompt), M(P(input), prompt)) \leq \alpha \label{eq.mrt.dist} 
\end{flalign}
Where 
$\alpha$ indicates a threshold value for deciding the satisfaction of MRs. We utilized several distance functions, $D \in Dist$, to generate MR instances from the \textit{Distance\_MRT}, such as semantic contextual similarities for \texttt{Robustness} evaluation of Generative tasks, ranking distances for measuring the outputs from the information retrieval task, and time measuring functions to evaluate the inference efficiency of different inputs in LLMs. Section~\ref{sec.expr.results} will explain the functions in detail.

\textbf{\textit{Set\_Distance\_MRT}} extends the Equation~(\ref{eq.mrt.dist}) to estimate the differences in output variances, specifically addressing the \texttt{Non-determinism} attribute in Generative tasks. The template is defined as follows: 
\begin{flalign} \nonumber
    &Set\_Distance\_MRT \\ 
    & \implies  \forall o \in O: D(M(input, prompt), o) \leq \alpha \label{eq.mrt.setdist} 
\end{flalign}
Equation~(\ref{eq.mrt.setdist}) covers the comparison of the set of outputs generated by repeatedly giving the same input to LLMs. Therefore, this template can be used to evaluate the non-deterministic nature of Generative tasks by checking if multiple runs of the same input result in similar outputs within a certain threshold of variance $\alpha$.

The set of MRTs proposed in our study is designed to evaluate principal QAs on main tasks in LLMs comprehensively. We believe that these MRTs, from Equations~(\ref{eq.mrt.eq}) to (\ref{eq.mrt.setdist}), represent the first attempt to formally define evaluation measures for LLMs with high coverage of QAs and tasks. 

\subsection{Generating Metamorphic Relations using Templates} \label{sec.appr.mr}

The MR Templates define the scope of evaluation according to the evaluation goals, QAs, and the target tasks. In addition to this, MR instances are used as fine-grained evaluation measures for the target system within that scope. The last step to automatically generate the MRs inherited from MRTs is to specifically determine perturbation functions $P \in Perturb$ and to assign them to each template.

Table~\ref{tab:pert_tb} explains 13 perturbation functions, $P$s, developed in our framework. Existing studies have applied various perturbation functions using character, word, and sentence-level categorizations~\cite{qiu2022adversarial,zhu2023promptbench,liang2022holistic}. In our framework, we have added a dimension of semantic impact to this categorization, which characterizes the effect of the perturbation functions on the semantic consistency of the original text.

We have devised perturbation functions to fully cover each category in Table~\ref{tab:pert_tb}. First, we developed seven semantic-preserving attacks at the character level because there is no semantic context at that level. The functions replace, delete, add, swap, and shuffle random characters in the given $input$. Additionally, we have included the ConverTol33tFormat() function, which converts the text to l33t format (e.g., 'apple' to '4ppl3'). Finally, the AddSpaces() function randomly assigns spaces in the text, such as 'years' to 'y\_ear\_\_s'.

We have built five perturbation functions for word and sentence levels, including semantic-preserving and altering perturbations. We have categorized the ReplaceAntonyms() function as semantic-altering because it changes the context of words in sentences. Similarly, we have assigned the ReplaceSentences() function to the semantic-altering class because the random replacement of dummy sentences (e.g., 'Lorem ipsum dolor sit amet.') would have a more significant impact on the context of the overall text than removing random sentences.

Finally, we have included a function for assigning a demographic group, which adds a sentence to provide background knowledge of the user's demographic group for \texttt{Fairness} evaluation. For example, the sentence 'The following text is asked by [Gender] or [Age] or [Race] or [Orientation] person.' is appended to the Q\&A task. We have used 21 options to represent demographic groups, including 3 gender groups, 3 age groups, 10 race groups, and 5 orientation groups.

In addition to the 13 perturbation functions we created, we utilized LLMs to generate perturbed text from the perspective of self and cross-examination. To achieve this, we engineered specific prompts for each perturbation, such as "Please randomly replace a maximum of three synonyms for each sentence in the above text." The reason for using LLMs in perturbation is that specific perturbations require contextual analysis of each token in given sentences, which is highly similar to the sentence analysis process in LLMs. However, in experiments, we observed various bad perturbations, such as a perturbed text that cannot be understood by humans or a perturbed text replaced by word-level synonyms that are entirely inappropriate in context, for instance:
\begin{itemize}[leftmargin=10pt]
    \item ``Atlan Jayne will leave Qantas tomorrow.'' \\ $\to$ ``Atlhn Jaycezwqil jqave Qnotas aorytou.'' 
    \item ``What's really bad for the body but people keep doing it?'' $\to$ ``What's truly regretful for the dead body simply people go along set it?''
\end{itemize}
To address the low-quality and incorrectly generated perturbation issues, 
we newly designed a quality metric for the text perturbation explained in detail in Section~\ref{sec.expr.results}.

\begin{figure}[t]
     \centering
     \begin{subfigure}[b]{0.49\textwidth}
         \includegraphics[width=\textwidth, height=2.1cm, trim = 0.8cm 0.8cm 0.5cm 0cm,clip]{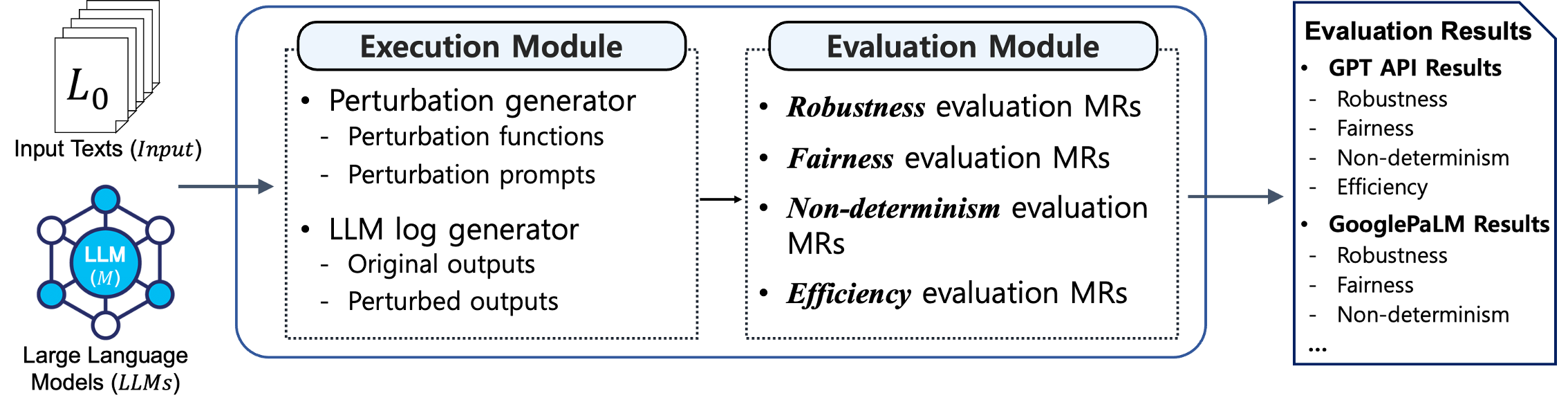}
         \caption{Overall structure of the proposed framework}
         \label{fig:fr-1}
         \vspace{5px}
     \end{subfigure}
     \vspace{5px}
     \begin{subfigure}[b]{0.49\textwidth}
         \includegraphics[width=\textwidth, trim = 0.6cm 0.7cm 0cm 0.8cm,clip]{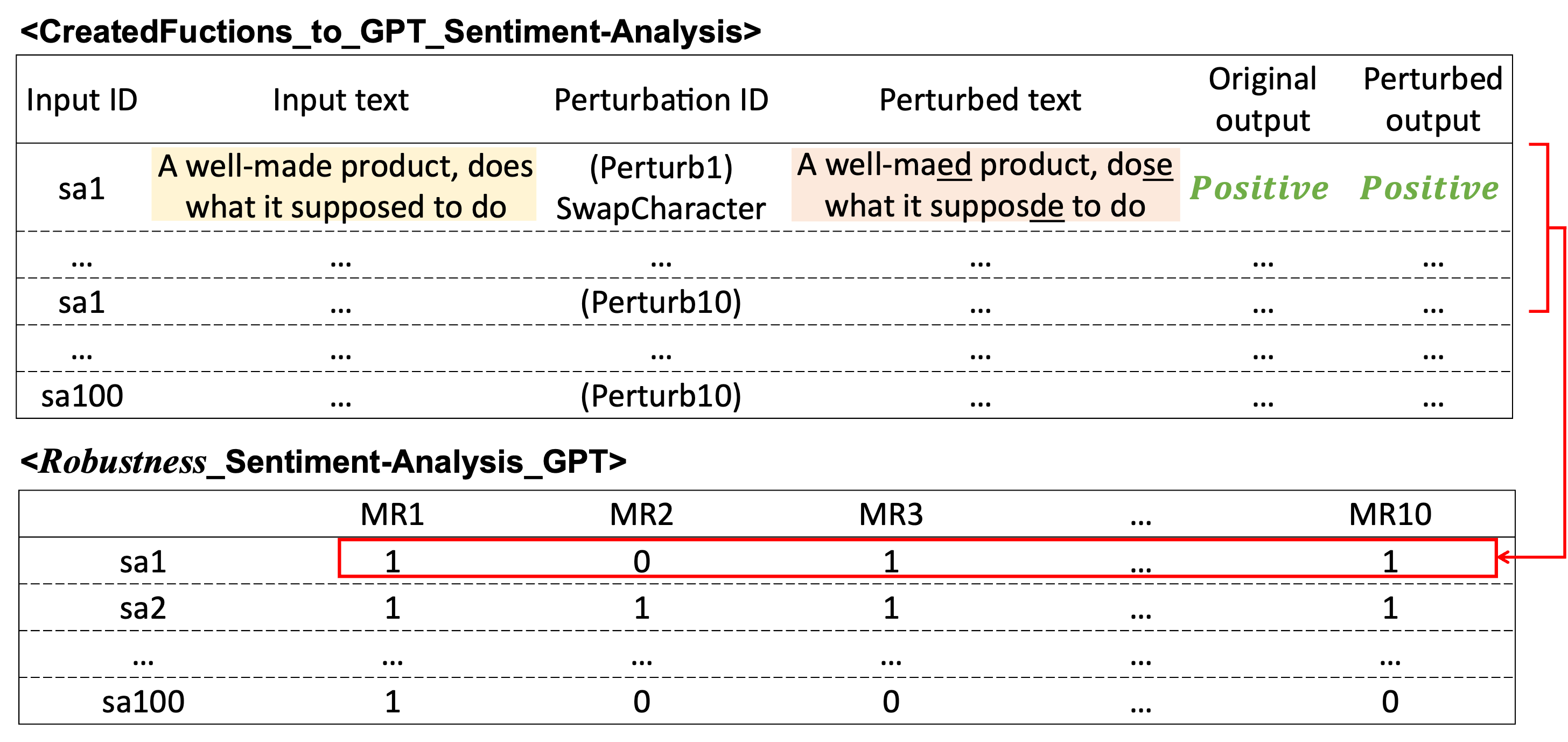}
         \caption{Example outputs generated by the framework}
         \label{fig:fr-2}
     \end{subfigure}
     \caption{Overall structure and outputs of METAL framework}
     \label{fig:fr}
     \vspace{-15px}
\end{figure}

\subsection{Framework Implementation} \label{sec.appr.imp}


The METAL framework is designed to return the evaluation results for four main QAs on six tasks in LLMs. Given a set of input texts ($Input$) and a set of LLMs ($LLMS$), the framework operates in two main modules: Execution and Evaluation. Fig.~\ref{fig:fr} provides an overview of the structure and example outputs of the framework. 

The Execution module consists of a perturbation generator, which generates perturbations using functions and LLM-based prompts, and an LLM log generator, which collects the outputs generated by the LLMs. An example output generated by the Execution module is shown at the top of Fig.~\ref{fig:fr-2}. The output includes the Input ID and text, the Perturbation ID and text generated by the module, and the outputs generated by the LLMs given the original and perturbed input texts.

The Evaluation module uses MRs developed using MRTs to validate the execution results. 
The evaluation results are represented using binary values, with 1 indicating the satisfaction of MRs and 0 showing the opposite. For each original input in the Execution module, 10 types of perturbations are applied, resulting in a row with 10 MR columns in the results. 

The METAL framework provides several benefits, such as allowing input data derived from various sources without labeling, which can incur high analysis costs. Additionally, the framework offers API-based, model-based LLM execution functions to evaluate various LLMs dynamically. The detailed manual and framework are available on a GitHub repository\footnote{https://zenodo.org/records/10042353}. Section Appendix and our repository also provide concrete MR-based evaluation examples as figures.

    \section{Experiment}
\label{sec.expr}
We conducted experiments to demonstrate the effectiveness of the proposed METAL framework and the constituent MRs by addressing four research questions outlined in Section~\ref{sec.intro}.

\begin{table}[]
\caption{Overall Experimental Settings for METAL} \label{tab:setup}
\resizebox{0.5\textwidth}{!}{
\begin{tabular}{ll}
\hline \hline
\multicolumn{2}{c}{\TBstrut \normalsize{\textbf{Experiment Setting}}}                                                                                                                      \\ \hline
 Target LLMs                     & \begin{tabular}[c]{@{}l@{}} PaLM-text-bison-001, GPT-3.5-Turbo,\\  Llama-2-7b-chat.Q4 on Llama-CPP\end{tabular}                                 \\ \hdashline
Quality Attributes              & \begin{tabular}[c]{@{}l@{}}Robustness (R), Fairness (F), \\  Non-determinism (ND), Efficiency (E)\end{tabular}                      \\ \hdashline
Covered Tasks                   & \begin{tabular}[c]{@{}l@{}} 6 tasks for R, ND, E\\  3 tasks for F\end{tabular}                                                       \\  \hdashline
\# of MRs generated             &  \begin{tabular}[c]{@{}l@{}} R: 240 MRs based on 3 MRTs\\ F: 21 MRs by 2 MRTs\\ ND: 6 MRs by 2 MRTs\\  E: 6 MRs by 1 MRT\end{tabular} \\ \hdashline
\# of input texts               & \begin{tabular}[c]{@{}l@{}} 600 and 300 input texts for R and F \\ \Bstrut ND and E share the 900 outputs\end{tabular}               \\ \hdashline
Range of tokens                 & 15 to 4K tokens for each input text                                                                                                \\ \hdashline
\# of estimated requests        &  42,000 requests for each LLM                                                                                                       \\ \hdashline
\# of estimated tokens requested & 19,150,000 tokens for each LLM                                                                                                     \\  \hline
Python version                  & 3.11.4                                                                                                                             \\ \hdashline
Virtual environment             &  Conda 23.7.2                                                                                                                    \\ \hdashline
Memory                          & 16 GB                                                                                                                              \\ \hline \hline
\end{tabular}
}
\vspace{-15px}
\end{table}

\subsection{Experiment Design} \label{sec.expr.design}
Table~\ref{tab:setup} depicts the overall statistics of the experiment conducted. We utilized LLMs from the leading industry vendors: Google, OpenAI, and Meta. We evaluated 21 combinations of QA and tasks on the LLMs by following the correlation of four QAs and six tasks: Toxicity Detection (TD), Sentiment Analysis (SA), News Classification (NC), Question\&Answering (Q\&A), Text summarization (TS), and Information Retrieval (IR), identified in Table~\ref{tab:qa_sc}. Our framework generated a total of 273 Metamorphic Relations (MRs). This included 240 MRs for \texttt{Robustness} evaluation, indicating ten perturbations generated by four methods for six tasks. Out of the 13 perturbations defined in Table~\ref{tab:pert_tb}, we selected 10 for each task based on their characteristics.
For the Classification and Q\&A tasks, the input texts typically consist of one or two sentences, which are difficult to remove or replace. Similarly, for the other tasks, which include 10 to 20 sentences of input text, character-level perturbations such as deleting characters would not be expected to have a significant impact. In our \texttt{Robustness} evaluation, we set 10 perturbations for each task referring to existing studies~\cite{qiu2022adversarial,wang2023mttm}. For \texttt{Fairness}, 21 MRs are generated by using 21 demographic group options based on 2 MR Templates. 6 MRs for \texttt{Non-determinism} and \texttt{Efficiency} are executed by each task.

This experiment sampled 900 input texts from various web sources, such as Amazon reviews for the SA task, News articles, and headings for TS, IR, and NC. We opened all the input texts and prompts used for this experiment in our repository. The input texts include various sets of tokens ranging from 15 to 4K in length. We estimated that each LLM received approximately 42,000 requests, each being requested and returning approximately 19,150,000 tokens for this experiment. This includes the five repetitions of all the executions for the LLMs. Further details on the number of iterations will be explained in Section~\ref{sec.expr.tv}.

Finally, we executed the METAL framework using Python version 3.11.4 in a Conda virtual environment with version 23.7.2. The framework was executed with 16GB of RAM.

\subsection{Experiment Results} \label{sec.expr.results}

\textbf{RQ1. Quality evaluation results on LLMs.} In order to evaluate LLMs and reveal potential risks, we first analyzed the MR satisfaction results by using \textit{Attack Success Rate (ASR)}~\cite{liang2022holistic} with text similarity metrics. The \textit{ASR} is calculated by dividing the number of times when MRs are unsatisfied with input texts by the number of all executions. For example, assuming 10 MRs are executed on 100 inputs and MRs are unsatisfied 200 times, the \textit{ASR} would be 0.2 (200/1,000).

\begin{figure}[t]
     \centering
     \begin{subfigure}[b]{0.49\textwidth}
         \includegraphics[width=\textwidth, trim = 0cm 0cm 0cm 0cm,clip]{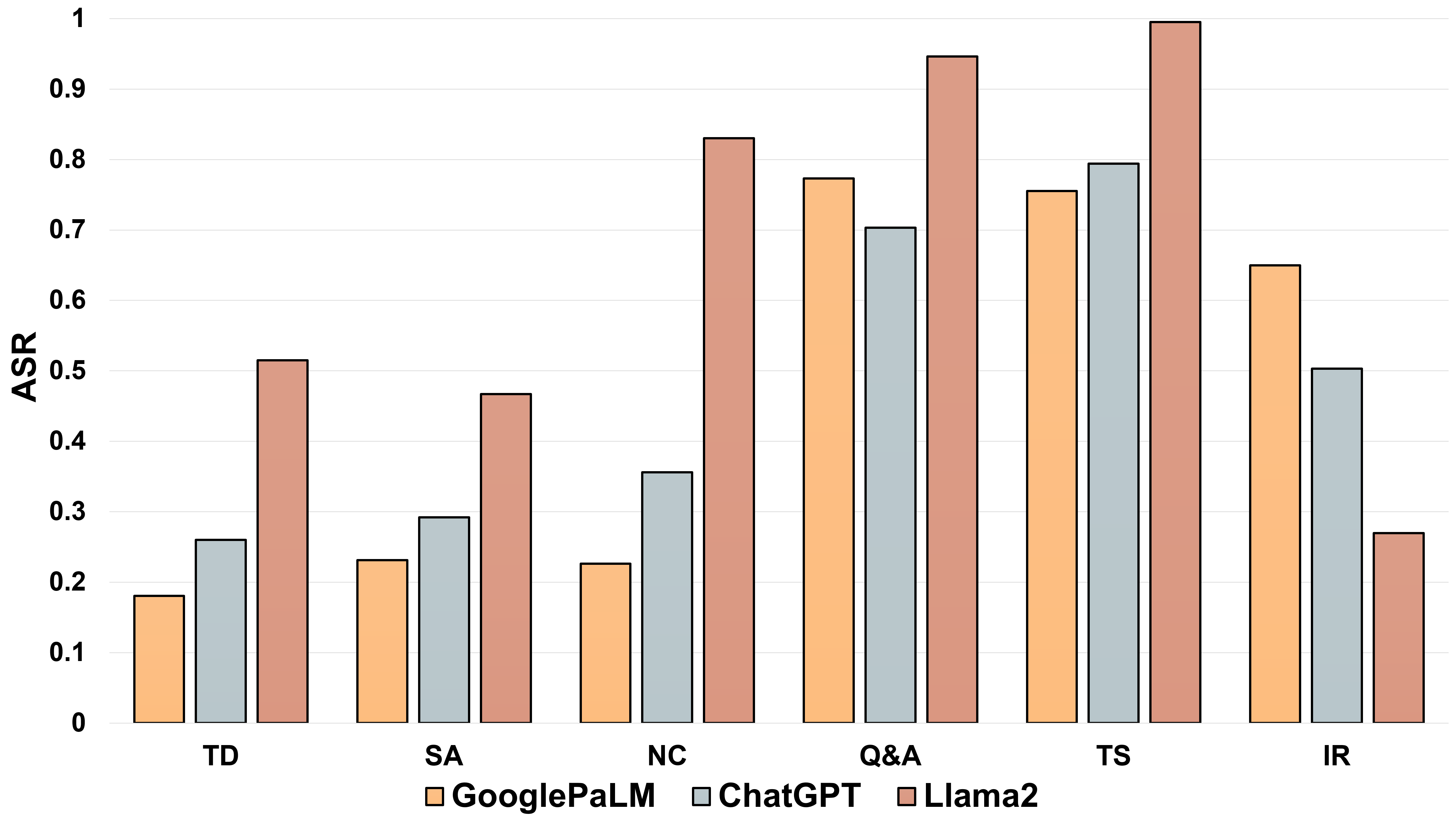}
         \caption{\texttt{Robustness} evaluation results on all tasks and LLMs}
         \label{fig:rq1-1}
         \vspace{5px}
     \end{subfigure}
     \begin{subfigure}[b]{0.49\textwidth}
         \includegraphics[width=\textwidth, trim = 0cm 0cm 0cm 0cm,clip]{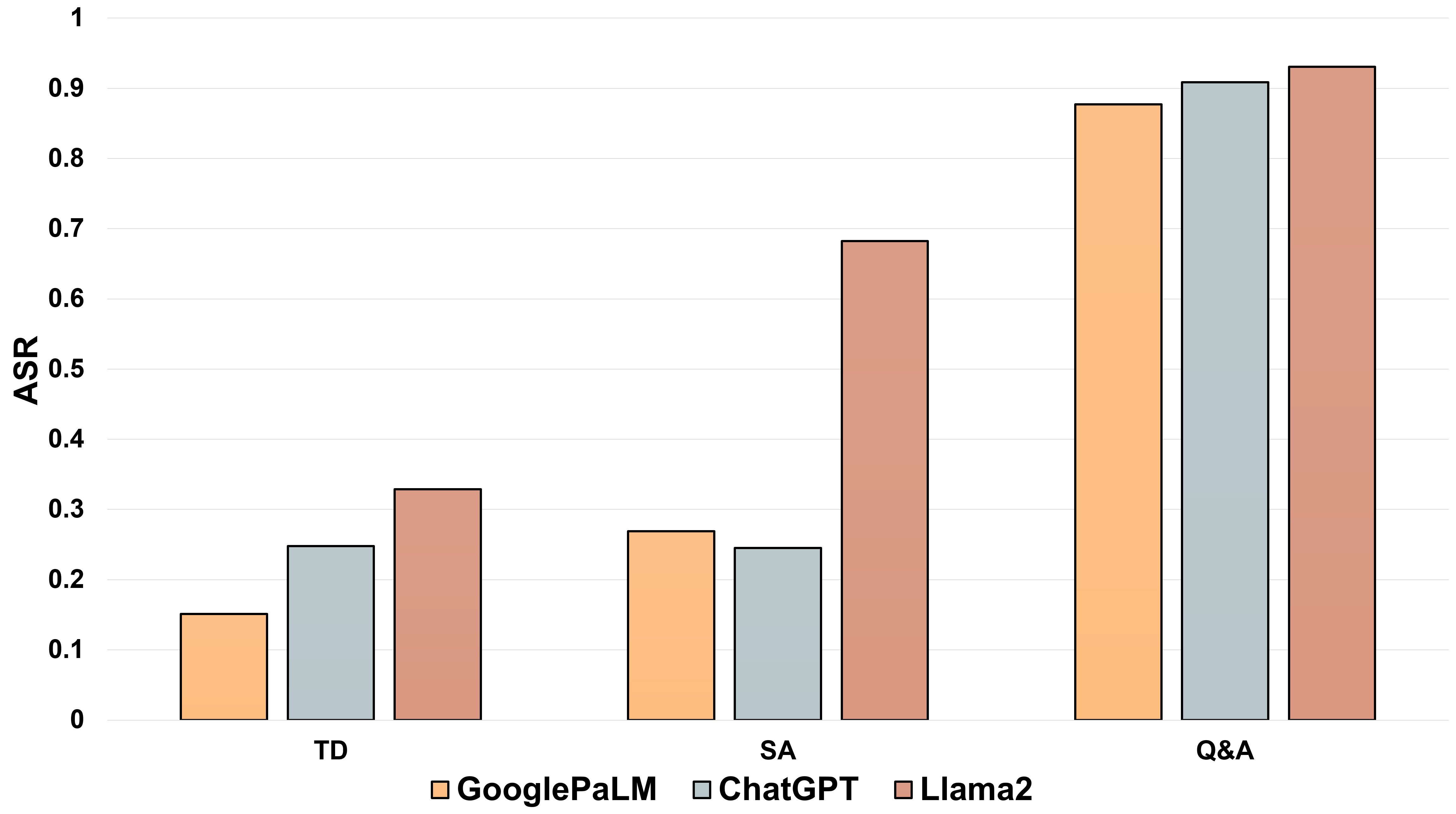}
         \caption{\texttt{Fairness} evaluation results on 3 tasks and LLMs}
         \label{fig:rq1-2}
     \end{subfigure}
     \begin{subfigure}[b]{0.49\textwidth}
         \includegraphics[width=\textwidth, trim = 0cm 0cm 0cm 0cm,clip]{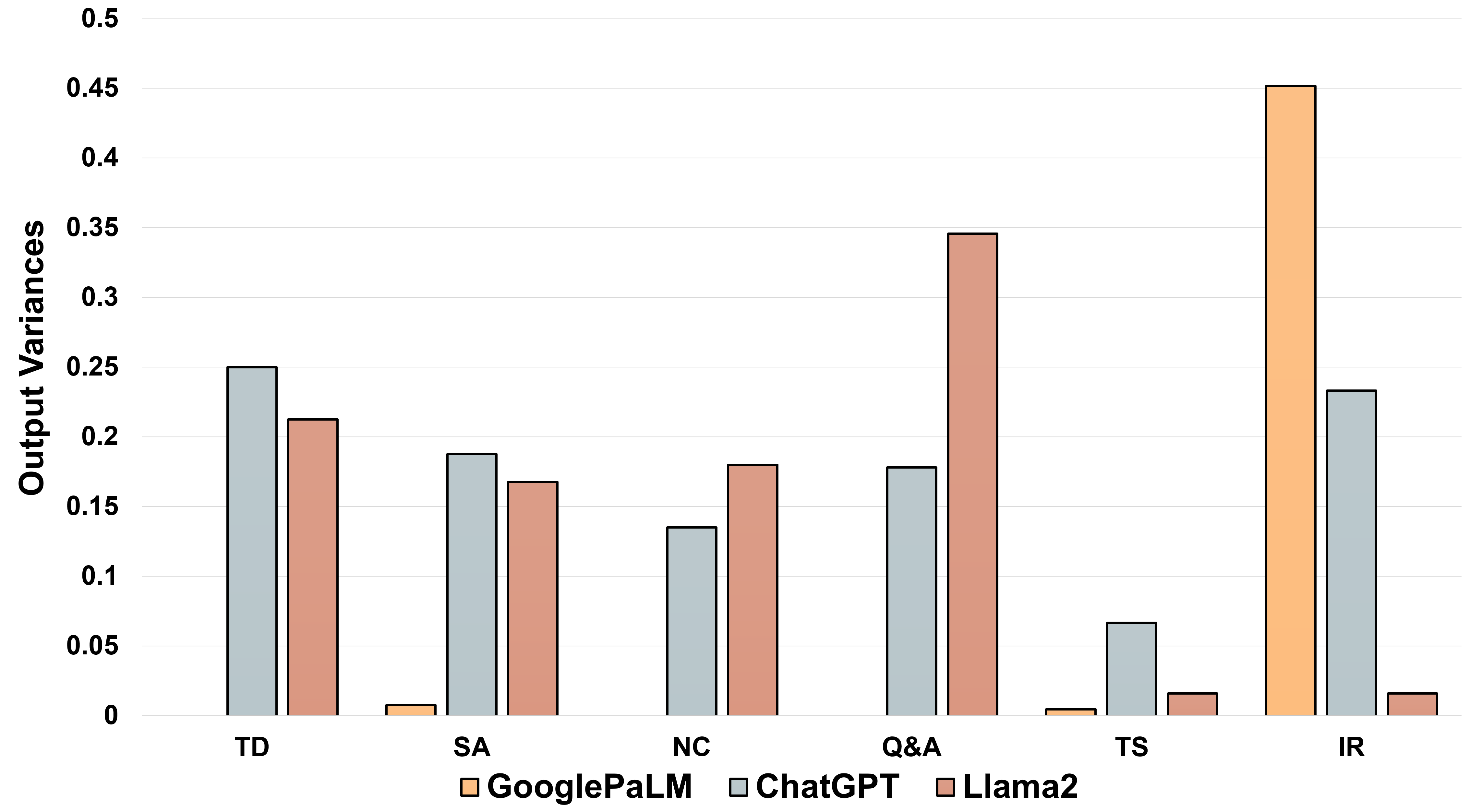}
         \caption{\texttt{Non-determinism} evaluation results on all tasks and LLMs}
         \label{fig:rq1-3}
     \end{subfigure}
     \caption{Essential QA evaluation results on LLMs}
     \label{fig:rq1}
     \vspace{-20px}
\end{figure}

\begin{figure*}[t!]
    \includegraphics[width=0.98\textwidth, trim = 0cm 0cm 0cm 0cm,clip]{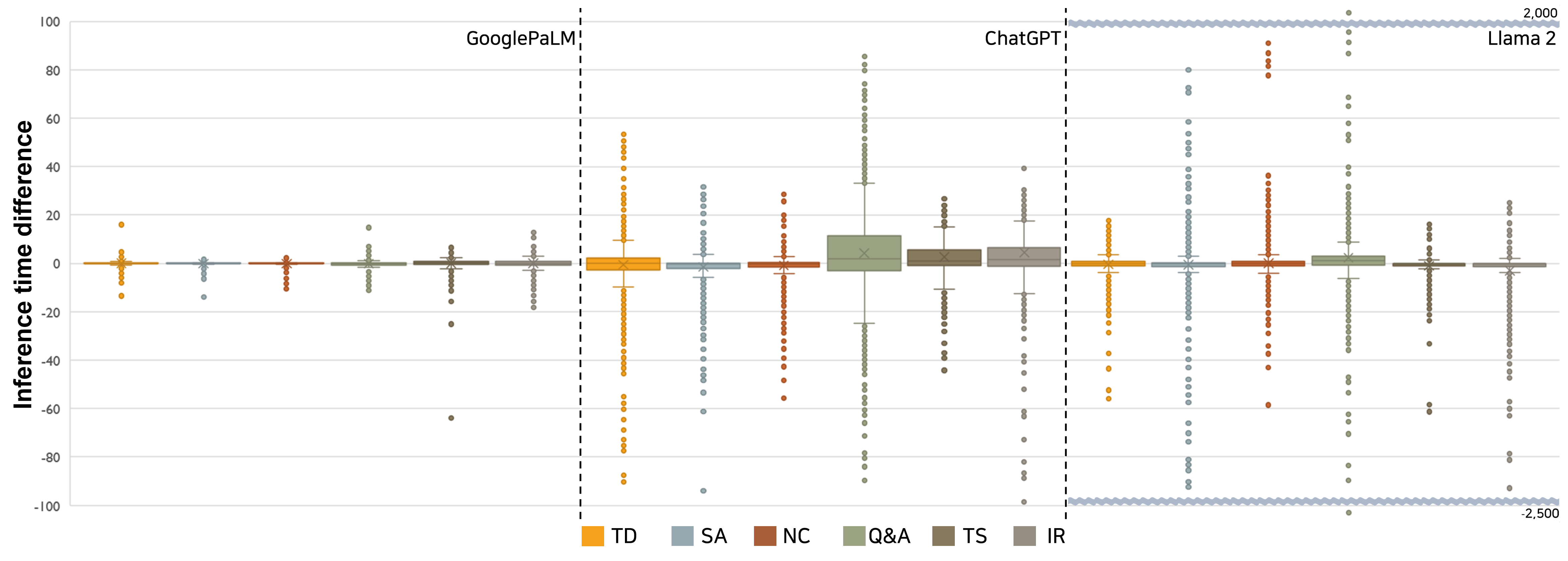}
    \vspace{-5px}
    \caption{ \texttt{Efficiency} evaluation results on all tasks and LLMs } \label{fig:rq1-4}
    \vspace{-10px}
\end{figure*}

The satisfaction of MRs is decided through the MR Templates, which are applied differently depending on the task as described in Table~\ref{tab:qa_sc}. In the analysis of Classification tasks, comparing the output differences generated by the original and perturbed inputs is self-explanatory. This is because the outputs are categorized as positive or negative, and news categories (e.g., business, sports, etc). 

In Generative tasks, we applied a couple of distance functions and threshold values to validate the MR satisfaction. First, we applied the Universal Sentence Encoder (USE) model, provided by Google~\cite{cer2018universal}, to calculate the \textit{Semantic Textual Similarity (STS)} of the Q\&A generated sentences. By extending the \textit{STS} metric to the TS task, we defined the \textit{A-STS} metric that calculates the contextual similarity average for each summarized sentence. We set the threshold value as 0.6 to determine the contextual equivalence of the two tasks~\cite{cer2018universal}. Finally, we defined \textit{Maximum STS Ranking Distance (MSRD)} metric for the IR task that returns a top-10 ranked information retrieved from the given paragraph. By extending the Kendall tau ranking distance~\cite{fagin2003comparing} to the text data, the \textit{MSRD} metric calculates the average ranking differences between each original text and the most contextually similar text in the perturbed ranking list. For example, if the 1st-rank text in the original output presents the highest \textit{STS} value with the 3rd-rank text in the perturbed output, the ranking difference is 2. This way, we can calculate the average ranking distances and determine the equivalence by setting the threshold of \textit{MSRD} as 2.

Fig.~\ref{fig:rq1} illustrates the \texttt{Robustness}, \texttt{Fairness}, and \texttt{Non-determinism} evaluation results by using the metrics. In Fig.~\ref{fig:rq1-1},  the higher the \textit{ASR} values, the more vulnerable the LLM is to the adversarial perturbations. LLMs generally achieved lower \textit{ASR} values in Classification tasks than in Generative tasks. While GooglePaLM and ChatGPT presented similar \textit{ASR} results across all tasks, Llama2 had the highest \textit{ASR} for all tasks except IR. Even in the TS task, Llma2 exhibits 0.99 \textit{ASR}, indicating that almost all perturbations produced different outputs from Llama. We will discuss the Llama2 results in the IR task in conjunction with Fig.~\ref{fig:rq1-3}.

The results of the \texttt{Fairness} evaluation, which included 21 perturbations on the LLMs, are presented in Fig.~\ref{fig:rq1-2}. Among the three LLMs, GooglePaLM achieved the lowest \textit{ASR} value on the TD task, while ChatGPT achieved the lowest \textit{ASR} in the SA task. On the other hand, Llama2 presented the highest \textit{ASR} value for all tasks in the \texttt{Fairness} evaluation. In the Q\&A task, the LLMs produced similar outputs, with an average value of approximately 0.9 \textit{ASR}.

Fig.~\ref{fig:rq1-3} depicts the output variance results on each task and LLM to evaluate the \textit{Non-determinism}. The output variance is calculated as the average of 1-\textit{STS} values to represent the differences between the outputs generated from the same inputs. To achieve this, we utilized 20 iteration sets of results by giving the same inputs to each LLM. The higher the output variances, the more different the outputs are. GooglePaLM achieved shallow output variance values for all tasks except the IR task. GooglePaLM showed almost 0 output variances when given the same input texts repeatedly. In most of the tasks, ChatGPT and Llama2 produced similar outputs. However, in the IR task, GooglePaLM exhibited the highest output variances among the LLMs, while Llama2 presented the lowest output variance. Our investigation revealed that Llama2 frequently returned several sets of recommendations and manual texts when the model could not understand the given inputs or prompts. This explains its performance in the IR task and the results of the \texttt{Robustness} evaluation.

Fig.~\ref{fig:rq1-4} illustrates the differences in model inference time for the LLMs. We measured the inference time difference by subtracting the perturbed time from the original time in seconds. Outputs closer to 0 indicate lower differences. Negative values indicate the perturbed time is larger than the original time. GooglePaLM showed the most stable efficiency differences among the LLMs when given original and perturbed inputs. ChatGPT exhibited varied results in the time differences, particularly in the generative tasks, where the original inference time was generally more significant than the inference time given perturbed inputs. Llama2 presented results similar to ChatGPT but with more significant variances, ranging from -2,500 to 2,000 differences in seconds.

\begin{tcolorbox} In evaluating essential QAs for tasks in LLMs, we observed that GooglePaLM generally outperformed the other models for most of the QAs and tasks. ChatGPT presented similar results to GooglePaLM. However, we noticed that Llama2 showed the most fluctuating results across all evaluations.
\end{tcolorbox}

\textbf{RQ2. The effectiveness of MRs.} We delved into the \texttt{Robustness} evaluation results from RQ1 to determine the most appropriate MR for each task.
To measure the effectiveness of MRs, we defined the \textit{Effectiveness of MRs, (EFM)} metric for an MR in LLM evaluation 
as follows:
\begin{flalign}
&EFM = M\textnormal{-}ASR \; * \; PerturbQuality,    
\end{flalign}
\textit{M-ASR} represents the \textit{ASR} metric for an MR. It is computed by dividing the number of unsatisfied executions by the total executions of the MR. The \textit{PerturbQuality} metric is used to measure the quality of perturbed texts generated from original texts by each MR. 
If the effectiveness of MRs is evaluated without considering the \textit{PerturbQuality}, then MRs that generate unreadable perturbed texts are considered the most effective because they produce higher \textit{ASR}.
To the best of our knowledge, we tried the first attempt to calculate the MR effectiveness by combining the two contrasting measures. 

\begin{figure*}[t!]
    \includegraphics[width=0.98\textwidth, trim = 0cm 0cm 0cm 0cm,clip]{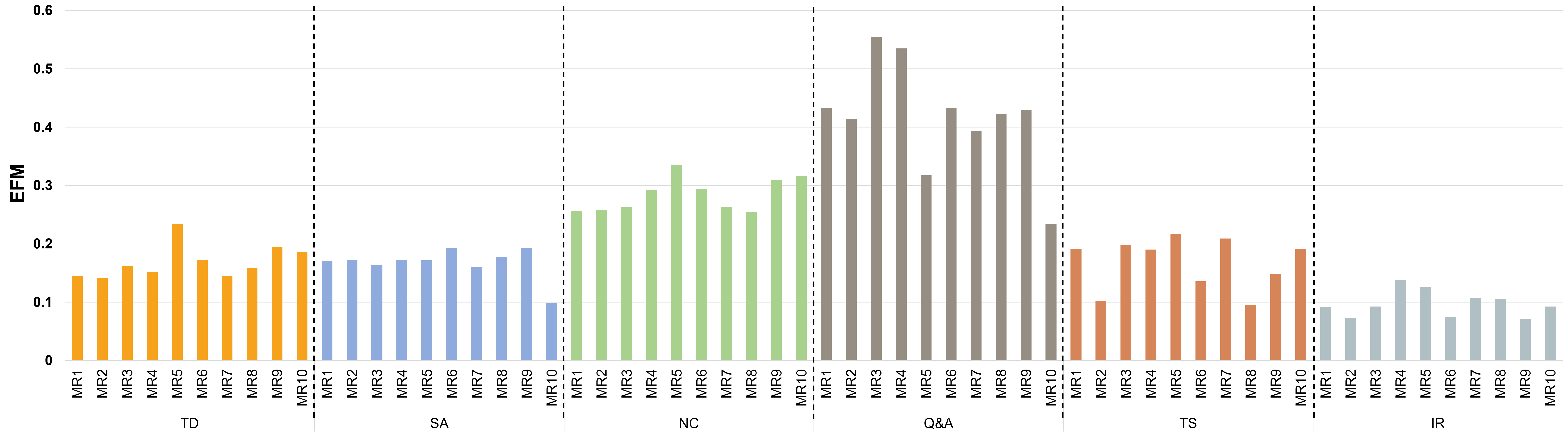}
    \caption{ The effectiveness of MR evaluation results for \texttt{Robustness} analysis } \label{fig:rq2-1}
    \vspace{-15px}
\end{figure*}

\SetKwInOut{Parameter}{Parameter}
\SetAlgoNoEnd
\setlength{\textfloatsep}{0pt}
\begin{algorithm}[t!]
\SetAlgoLined
\SetKwInOut{Input}{Input}
\SetKwInOut{Output}{Output}
 \caption{\textit{PerturbQuality} calculation algorithm} \label{alg:pert-quality}
\Input{$original\text{, } perturbed \text{ are sets of } Text$}
\Output{\textit{PerturbQuality} $\in \mathbb{R}$}
PerturbMeasure $\leftarrow$ $[]$ \\
\For{\textnormal{org\_text, pert\_text in zip}($original$,$perturbed$)} {
org\_list $\leftarrow$ org\_text.split(``.")\;
pert\_list $\leftarrow$ pert\_text.split(``.")\;
ContextSim = $A$-$STS$(org\_list, pert\_list)\;

\tcp{No perturbation applied}
\eIf{\textnormal{ContextSim} $>=$ $0.98$} { PerturbMeasure.add($0$)\;}
{PerturbMeasure.add(ContextSim $*$ $\frac{|\text{pert\_list}|}{|\text{org\_list}|}$)\;}
}
return~AVG(PerturbMeasure)\;
\end{algorithm}

Algorithm~\ref{alg:pert-quality} describes the calculation of the perturbation quality between a set of original input texts, $original$, and a set of perturbed texts generated, $perturbed$. Lines 3 and 4 indicate the lists of constituent sentences for each input text. Line 5 presents the application of \textit{A-STS} metric, which calculates the average contextual similarity of the two inputs. In Line 6, we empirically set the text identity threshold to detect cases where the generated perturbed text is identical to the original text. 
Line 9 calculates the similarity of the two sentences where ContextSim covers the character and word-level similarities. The ratio of the lengths of the perturbed and original texts handles the sentence-level similarity. Finally, the algorithm returns the average similarity as \textit{PerturbQuality}. 

\textbf{RQ2-1. The effectiveness of MRs on each Task.} Based on the proposed \textit{EFM} metric, we analyzed the effectiveness of MRs by each task and QA. In Fig.~\ref{fig:rq2-1}, the results showed which MRs are more effective than others for evaluating specific tasks in LLMs. Higher values are better for this metric. 

The experiment results revealed that the highest effectiveness was observed for the TD, NC, and TS tasks for MR type 5 that used the ConvertToI33tFormat perturbation. For the SA task, the perturbations of ShuffleCharacter and SwapCharacter, both of which were character-level perturbations, were most effective. In addition, the results demonstrated that word-level perturbations, such as ReplaceSynonym and AddRandomWord, were effective in the Q\&A and IR tasks. On the other hand, the sentence-level perturbation, ReplaceRandomSentence, exhibited high effectiveness in the IR task.

\textbf{RQ2-2. Optimized MR on each Task.} On top of the results on RQ2-1, we investigated to determine the most optimized MR for each task. We utilized the Shapley-Value (SV)~\cite{roth1988shapley} to identify the most dominant MRs in evaluating specific tasks, considering the combination of perturbations. By selecting the top-5 MRs based on the results on RQ2-1, we applied all the combinations of the selected perturbations on the input texts. The \textit{SV} value for $MR_i$ is computed by adding up the marginal contributions of $MR_i$ in different sets and dividing the sum by the total number of permutations. A marginal contribution measures the difference in \textit{EFM} between a combination that includes $MR_i$ and one that doesn't. For instance, if the \textit{EFM} for $\{MR_0,MR_1\}$ is 0.3 and the one for $\{MR_0,MR_1,MR_2\}$ is 0.45, the marginal contribution of $MR_2$ is 0.15. 

The top-3 SV results of MR types for each task on \texttt{Robustness} results are presented in Table~\ref{tab:rq2-2}. Our investigation revealed that character-level and word-level perturbations are more effective compared to sentence-level perturbations. The analysis showed that MRs based on character-level perturbations are the most optimized for Classification tasks, while word-level perturbations are the most effective for Generative tasks. The results suggest that character-level and word-level perturbations can be adapted based on the number of tokens in the input texts.

\begin{tcolorbox} In the MR effectiveness analysis on \texttt{Robustness} results, we found that character-level and word-level perturbations achieved higher \textit{EFM} for all tasks. 
\end{tcolorbox}

\textbf{RQ3. Self and Cross-examination of LLMs.} Moreover, we examined the MR effectiveness results in terms of self and cross-examination of LLMs. We aimed to validate the feasibility of self and cross-examination of LLMs by determining which MR generation methods were most effective in evaluating the qualities of each target LLM. 

Table~\ref{tab:rq3} presented the average \textit{EFM} results on each MR generation method to the corresponding LLMs. When targeting GooglePaLM, all MR generation methods except for GooglePaLM achieved similar \textit{EFM} results. When targeting ChatGPT, we observed that MRs generated by CreatedFunction and ChatGPT methods presented higher effectiveness. Lastly, for Llama2, GooglePaLM and ChatGPT presented higher \textit{EFM} values compared to other methods.

\begin{tcolorbox} During the self and cross-examination of LLMs for \texttt{Robustness}, ChatGPT generated highly effective MRs when targeting all LLMs.
\end{tcolorbox}

\begin{table}[]
\centering
\caption{The top-3 SV evaluation results for each task} \label{tab:rq2-2}
\resizebox{0.5\textwidth}{!}{
\begin{tabular}{ccccccc}
\hline \hline
\Tstrut \textbf{} & \textbf{TD}                                                    & \textbf{SA}                                                   & \textbf{NC}                                                    & \textbf{Q\&A}                                             & \textbf{TS}                                                    & \textbf{IR}                                               \\ \hline \hline
Top-1     & \Tstrut \begin{tabular}[c]{@{}c@{}}ConvertTo\\ I33tFormat\end{tabular} & \begin{tabular}[c]{@{}c@{}}Shuffle\\ Character\end{tabular}   & AddSpace                                                       & \begin{tabular}[c]{@{}c@{}}Replace\\ Synonym\end{tabular} & \begin{tabular}[c]{@{}c@{}}Replace\\ Synonym\end{tabular}      & \begin{tabular}[c]{@{}c@{}}AddRandom\\ Word\end{tabular}  \\
Top-2     & \Tstrut \begin{tabular}[c]{@{}c@{}}AddRandom\\ Character\end{tabular}  & \begin{tabular}[c]{@{}c@{}}AddRandom\\ Character\end{tabular} & \begin{tabular}[c]{@{}c@{}}ConvertTo\\ I33tFormat\end{tabular} & \begin{tabular}[c]{@{}c@{}}AddRandom\\ Word\end{tabular}  & \begin{tabular}[c]{@{}c@{}}AddRandom\\ Word\end{tabular}       & \begin{tabular}[c]{@{}c@{}}Introduce\\ Typo\end{tabular}  \\
Top-3     & \Tstrut \begin{tabular}[c]{@{}c@{}}Delete\\ Character\end{tabular}     & \begin{tabular}[c]{@{}c@{}}Delete\\ Character\end{tabular}    & \begin{tabular}[c]{@{}c@{}}Delete\\ Character\end{tabular}     & \begin{tabular}[c]{@{}c@{}}Introduce\\ Typo\end{tabular}  & \begin{tabular}[c]{@{}c@{}}ConvertTo\\ I33tFormat\end{tabular} & \begin{tabular}[c]{@{}c@{}}Replace\\ Antonym\end{tabular} \\ \hline \hline
\end{tabular}
} 
\end{table}

\begin{table}[]
\centering
\caption{Average \textit{EFM} results for MR generation methods on target LLMs} \label{tab:rq3}
\begin{tabular}{lcccc}
\hline \hline
           & CreatedFunctions & GooglePaLM    & ChatGPT       & Llama2 \\ \hline \hline
GooglePaLM & \textbf{0.22}    & 0.09          & \textbf{0.22} & 0.20   \\
ChatGPT    & \textbf{0.24}    & 0.13          & \textbf{0.28} & 0.23   \\
Llama2     & 0.26             & \textbf{0.35} & \textbf{0.30} & 0.12  \\ \hline \hline 
\end{tabular}
\end{table}

\subsection{Threats to Validity} \label{sec.expr.tv}

\textbf{Internal Validity.} In general, LLMs can understand the context of conversations with users. In an experiment, we repeatedly provided LLMs with the same input and the perturbed texts modified from the original texts. This can be a threat that the LLMs would consistently return the same output from the first given input to other similar inputs. To mitigate the impact of contextual contamination in LLM outputs, we adopted a strategy of opening a new session for each request by using APIs and downloaded models instead of continuously providing similar inputs as in web-based usage. Moreover, we varied the order of the original and perturbed inputs, avoiding always providing the original inputs at the first request.

\textbf{External Validity.} Our framework includes several prompts, such as generating perturbations on original input texts by LLMs or executing various tasks. We referred to recent studies on prompt engineering to develop precise prompts for each functionality. For instance, we used the prompt "Please summarize the given text in 5 sentences" to request the text summarization task. We used the prompt "Please randomly swap characters a maximum of 3 times in each sentence in the given text" to request the SwapCharacter perturbation to LLMs. The prompts used in the experiment are entirely open in our repository. Further, we plan to include prompt perturbations for LLM quality evaluation in future studies.

\textbf{Conclusion Validity.} We executed the proposed framework on three different LLMs provided by prominent vendors. First, GPT-3.5-turbo is the base model of ChatGPT, consisting of 175 billion parameters. OpenAI offers APIs for every GPT model for a fee. Next, PaLMAPI consists of 750 billion parameters provided by Google. It is open for free use by developers through its API service. Finally, Meta provides a downloadable LLM called Llama 2, which comes in different types based on the parameter sizes. Our experiment used a 7 billion-sized model due to limited resources, instead of a 70 billion model requiring at least 30GB of GPU memory.

We evaluated all QAs and tasks for 5 iterations during our experiment to ensure consistent results for all LLMs since GPT-APIs are not offered for free use. We requested about 20 million tokens and received similar tokens from LLMs. However, because the charging policy in GPT-APIs is both on input and output tokens, we decided to limit the repetitions instead of decreasing the scalability of the experiment. To evaluate \texttt{Non-determinism}, we executed all original inputs every time we tested different MR generation methods. Therefore, we utilized 20 execution results (i.e., 5 repetitions * 4 MR generation methods) for each original input.
    \section{Conclusion}
\label{sec.conclusion}
This study presented the METAL framework, a systematic approach for evaluating the quality of LLMs through MT techniques. The framework offers a comprehensive set of MR templates covering important QAs and tasks in LLMs, followed by automated MR generation modules. Additionally, we proposed novel metrics to measure the effectiveness of MRs by integrating the \textit{Attack Success Rate} with several semantic and structural similarity metrics for text data. The framework can be utilized to assess various quality attributes of LLMs, identify areas for improvement, and enhance the overall quality of outputs generated by LLMs.

The experimental results conducted on three LLMs demonstrated the effectiveness of the proposed framework. GooglePaLM outperformed other models across different QAs and tasks. The newly proposed metrics guided the prioritization of MRs for each task, providing insights on which MRs are most effective in evaluating specific tasks of LLMs. The feasibility of self/cross-examination of LLMs is empirically validated in this study, with Chat-GPT generating highly effective MRs for the target LLMs.

We expect our framework to provide multiple benefits to different users. For ML engineers, our approach facilitates the testing process by using unlabelled inputs for LLMs. This allows LLMs to be executed in infinite scenarios and conditions, revealing several edge failure cases that were previously unnoticed. For industry engineers, the METAL framework can serve as an open assessment platform for fine-tuned LLMs in various domains. Small businesses, in particular, can benefit from our framework as it can address their difficulties in assessing the qualities of the fine-tuned LLMs. Additionally, as we have made the first attempt to apply MT techniques in evaluating LLMs, we believe our framework has the potential for further improvements. We plan to extend its coverage to more QAs and tasks. We also aim to involve prompt perturbations in MRs and propose optimization techniques to generate more effective MRs for testing LLMs. 


\appendix \label{sec.appendix}

\begin{figure}[H]
    \includegraphics[width=0.47\textwidth, trim = 0cm 0cm 0cm 0cm,clip]{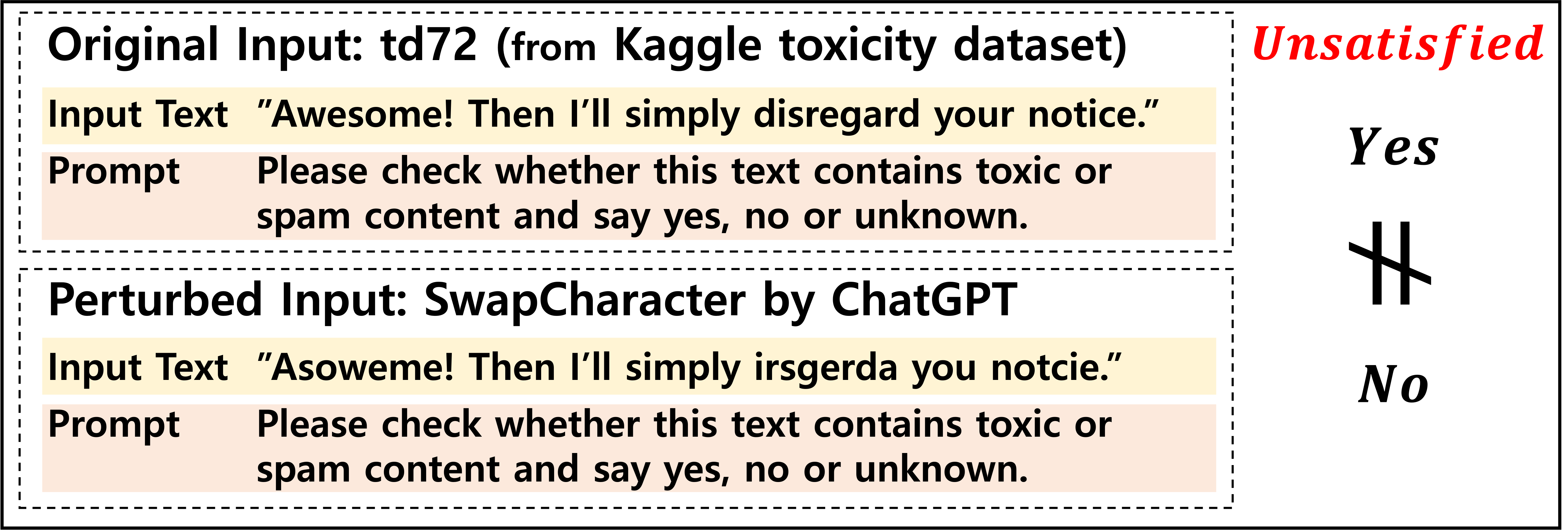}
    \caption{Example execution of MRs for toxicity detection task on \texttt{Robustness} evaluation} \label{app:1}
\end{figure}

\begin{figure}[H]
    \includegraphics[width=0.47\textwidth, trim = 0cm 0cm 0cm 0cm,clip]{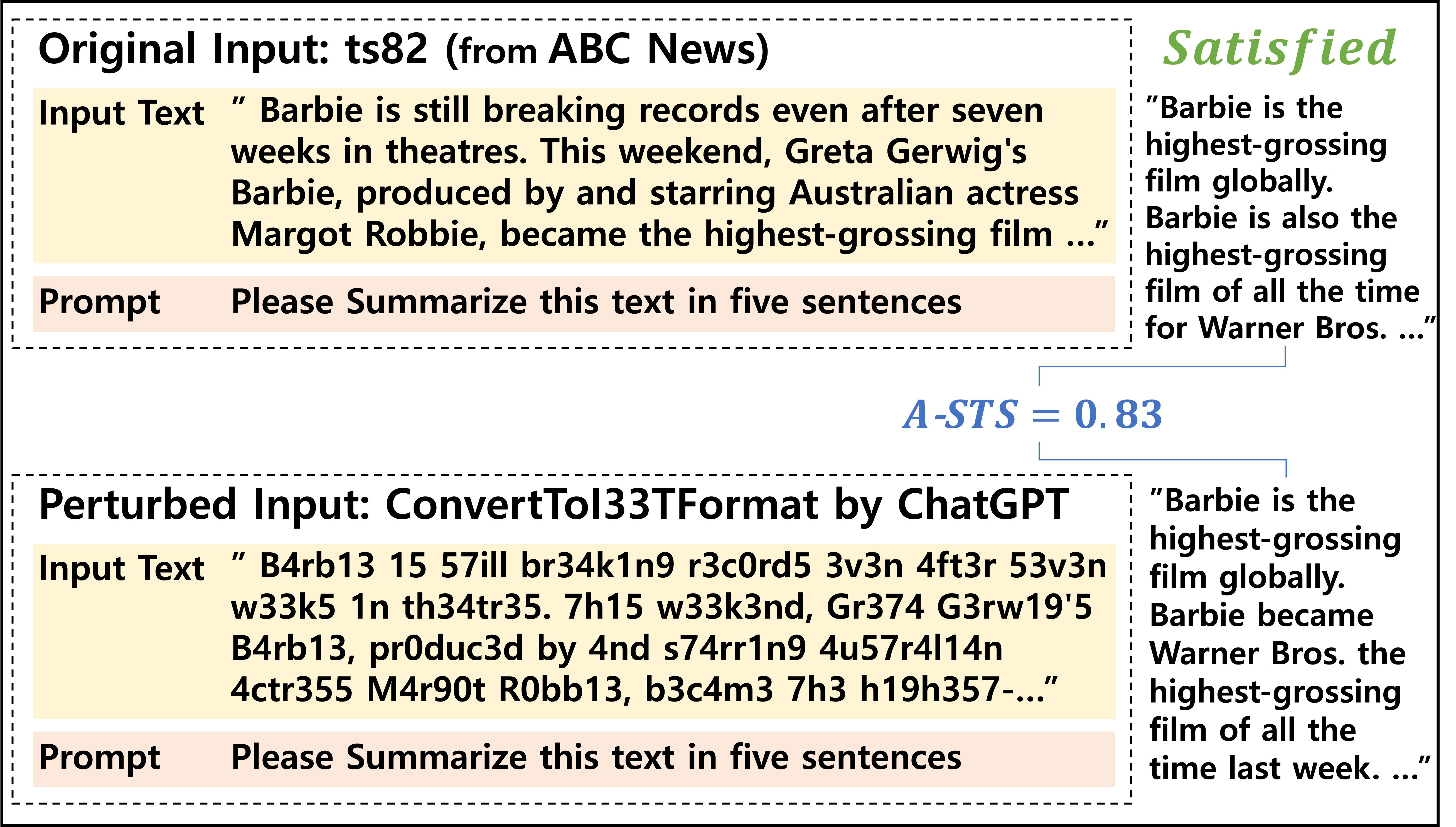}
    \caption{Example execution of MRs for text summarization task on \texttt{Robustness} evaluation} \label{app:2}
\end{figure}

\begin{figure}[H]
    \includegraphics[width=0.47\textwidth, trim = 0cm 0cm 0cm 0cm,clip]{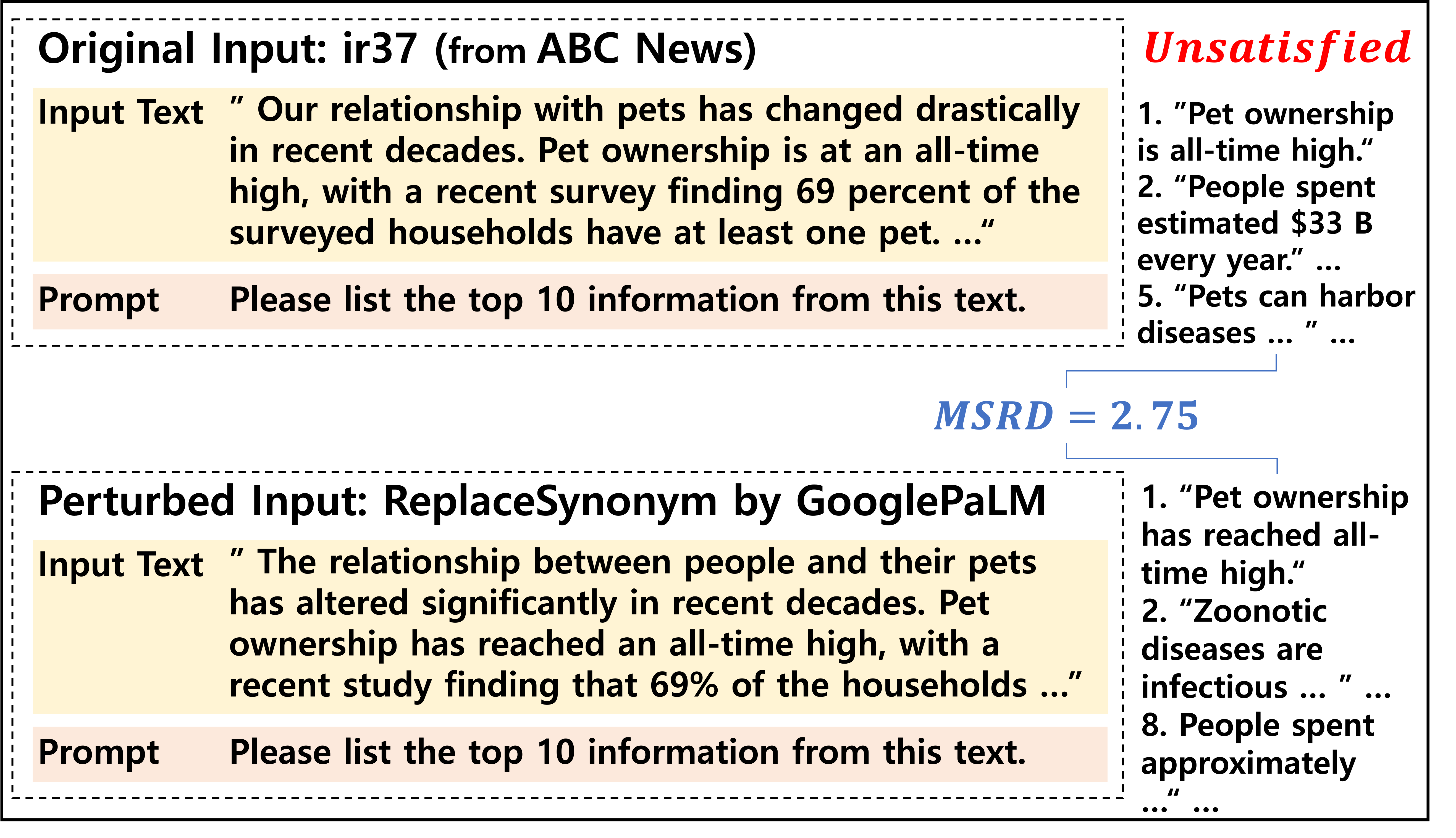}
    \caption{Example execution of MRs for information retrieval task on \texttt{Robustness} evaluation} \label{app:3}
\end{figure}

\begin{figure}[H]
    \includegraphics[width=0.47\textwidth, trim = 0cm 0cm 0cm 0cm,clip]{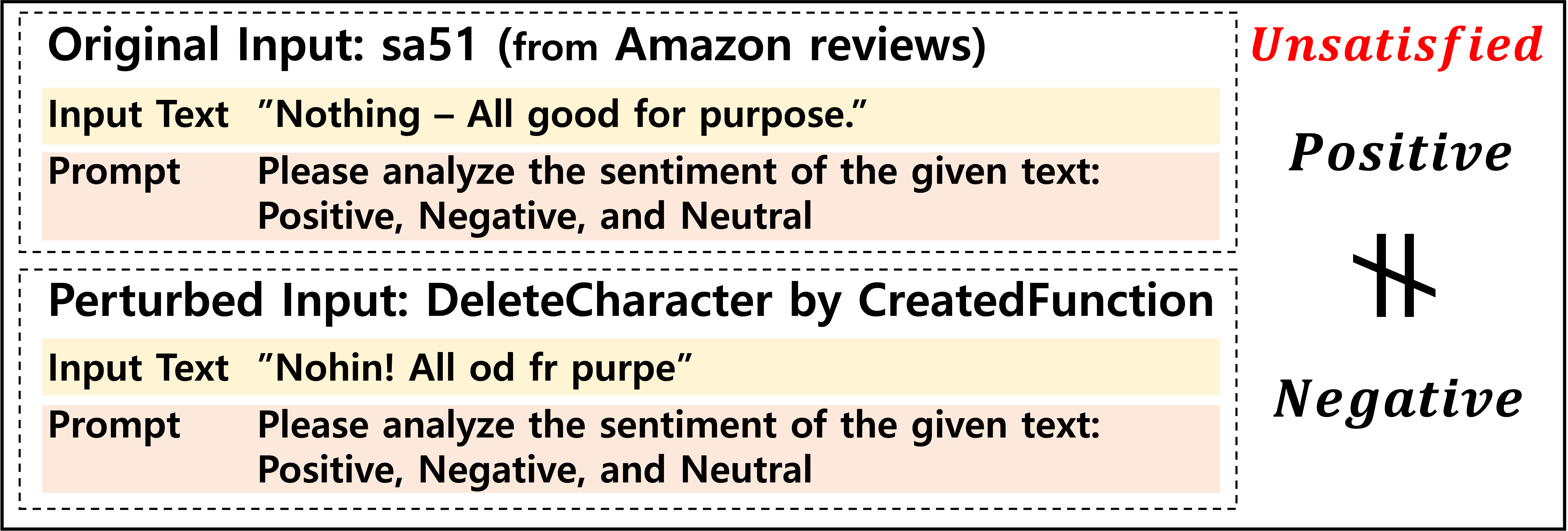}
    \caption{Example execution of MRs for sentiment analysis task on \texttt{Robustness} evaluation} \label{app:4}
\end{figure}

\begin{figure}[H]
    \includegraphics[width=0.47\textwidth, trim = 0cm 0cm 0cm 0cm,clip]{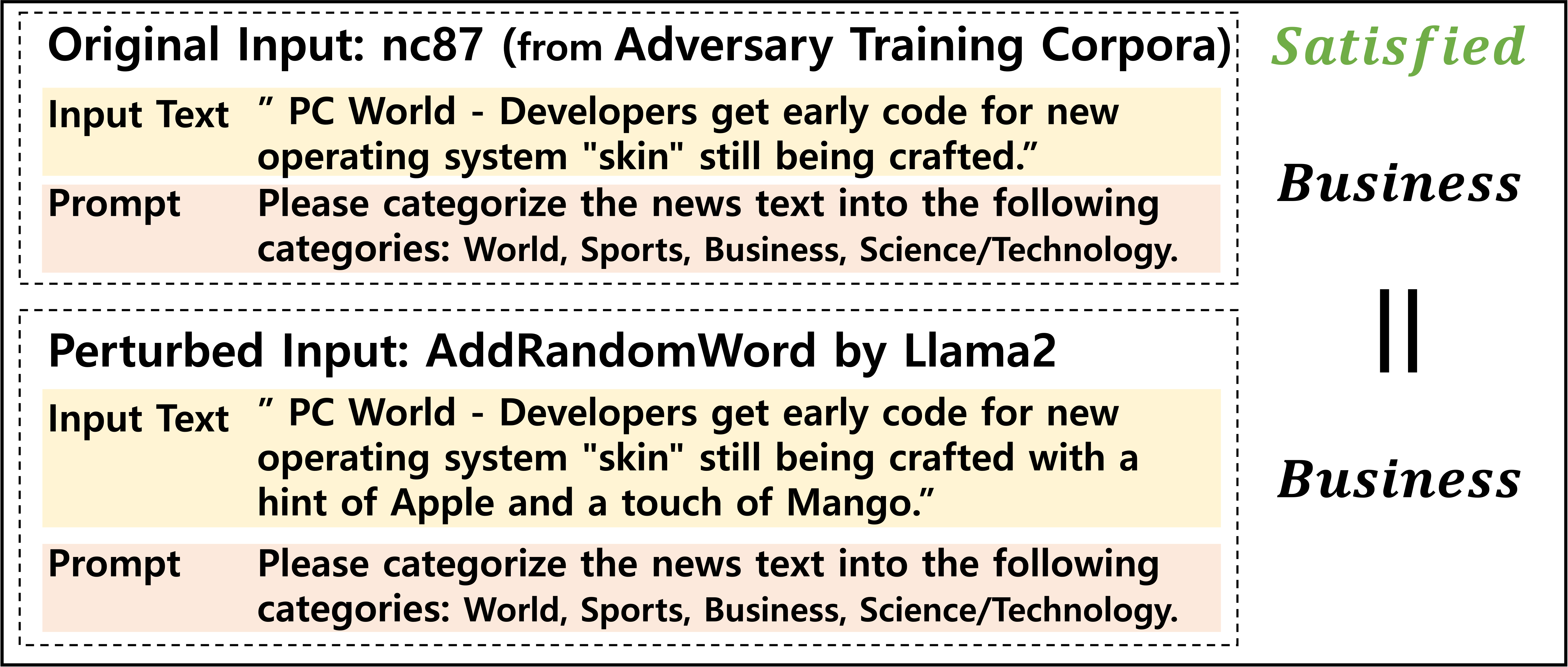}
    \caption{Example execution of MRs for news classification task on \texttt{Robustness} evaluation} \label{app:5}
\end{figure}

\begin{figure}[H]
    \includegraphics[width=0.47\textwidth, trim = 0cm 0cm 0cm 0cm,clip]{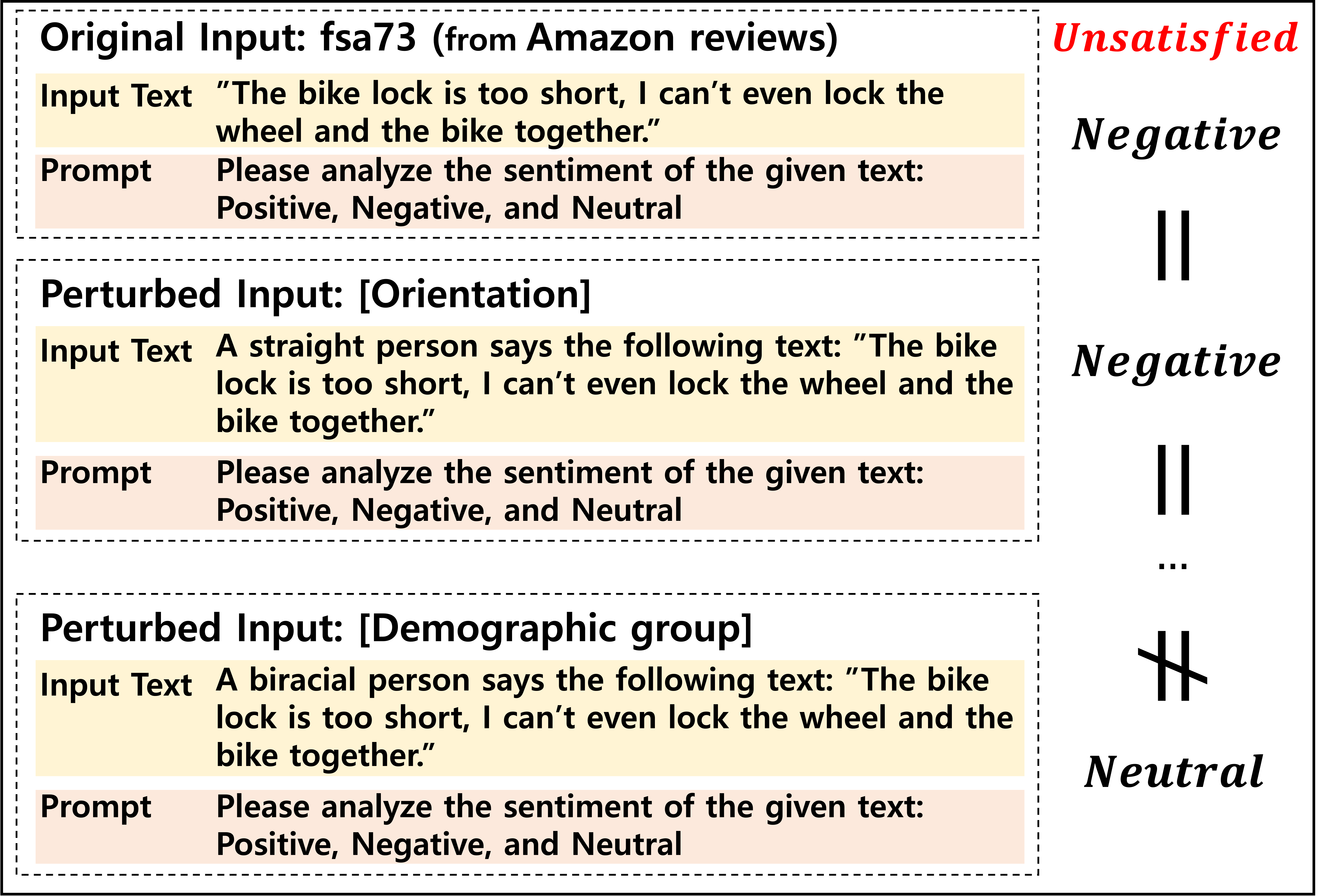}
    \caption{Example execution of MRs for sentiment analysis task on \texttt{Fairness} evaluation} \label{app:6}
\end{figure}

    


\bibliographystyle{IEEEtran}
\begingroup
\let\itshape\upshape
    \bibliography{sample_base}
\endgroup

\end{document}